\documentclass[11pt]{article}
\usepackage[utf8]{inputenc}
\usepackage[left=1.2in,top=.8in,right=1.2in,nohead,bottom=1in]{geometry}

\usepackage{amssymb}
\usepackage{amsthm}
\usepackage{color,comment}

\usepackage{commath}
\usepackage{enumerate,setspace}
  
\newtheorem{theorem}{Theorem}
\newtheorem{remark}{Remark}
\usepackage{amsmath}
\usepackage{natbib}
\usepackage{graphicx}
\allowdisplaybreaks

\graphicspath{ {./} }

\newcommand{\Var}{\text{Var}}

\newcommand{\bx}{\mathbf{x}}
\newcommand{\btheta}{\boldsymbol\theta}

\title{A principled stopping rule for importance sampling}

\author{
  Medha Agarwal \\
  Department of Statistics\\
  University of Washington\\
  \texttt{medhaaga@uw.edu} \and
 Dootika Vats\\
  Deptartment of Mathematics and Statistics\\
  IIT Kanpur\\
  \texttt{dootika@iitk.ac.in} \and
  V\'{i}ctor Elvira  \\
  School of Mathematics \\ 
  University of Edinburgh\\
  \texttt{victor.elvira@ed.ac.uk} 
}

\date{}

\begin{document}

\maketitle  
 \onehalfspacing    
 
\begin{abstract}
Importance sampling (IS) is a Monte Carlo technique that relies on weighted samples, simulated from a proposal distribution, to estimate intractable integrals.  The quality of the estimators improves with the number of samples. However, for achieving a desired quality of estimation, the  required number of samples is unknown and depends on the quantity of interest, the estimator, and the chosen proposal. We present a sequential stopping rule that terminates simulation when the overall variability in estimation is relatively small. The proposed methodology closely connects to the idea of an effective sample size in IS and overcomes crucial shortcomings of existing metrics, e.g., it acknowledges multivariate estimation problems.  Our stopping rule retains asymptotic guarantees and provides users a clear guideline on when to stop the simulation in IS. 
\end{abstract} 

\section{Introduction}
 
In a wide variety of applications, a key problem of interest is the estimation of intractable integrals through Monte Carlo techniques. Specifically, let ${\pi}(x)$ be a target distribution on $\mathcal{X} \subseteq \mathbb{R}^s$ with associated density function also denoted by $\pi$. Suppose $h: \mathcal{X} \to \mathbb{R}^p$ and interest is in estimating
\begin{equation}
    \mu_h := \int_{\mathcal{X}} h(x) {\pi}(dx)  < \infty\,.
\end{equation}
Under independent and identically distributed (iid) sampling from
$\pi$, the vanilla Monte Carlo estimator for $\mu_h$ is \vspace{-.2cm}
\begin{equation}
\bar{\mu}_h := \dfrac{1}{n}\sum_{i=1}^{n}h(Y_i), \quad \text{ where } Y_i \,\overset{\text{iid}}{\sim}\, {\pi}  \text{ for } i = 1, \dots, n\,.
\end{equation}

While iid sampling from the target density is often desirable, it may not always be feasible either due to the computational burden or  the inefficiency of the resulting estimators \citep[see][]{denny2001introduction}.  In such cases, practitioners may resort to other Monte Carlo methods such as  Markov chain Monte Carlo (MCMC) \citep{robert2013monte,ekvall2014markov} or importance sampling (IS) \citep{kahn1950random1, kahn1950random2,elvira2021advances}.
 
 Importance sampling is a popular Monte Carlo technique often used for variance reduction. In IS, samples from a proxy \textit{proposal} distribution are simulated and weighted averages of desired functions are calculated.  Specifically, let $q(x)$ denote the density of a chosen proposal distribution and for a known or unknown normalizing constant $Z$, let $\widetilde{\pi}(x) = {Z{\pi}(x)}$ be such that it can be evaluated (often called \emph{unnormalized target}) for every $x \in \mathcal{X}$.  Random samples, $X_1, \dots, X_n$ drawn from the distribution with density $q$ are assigned importance weights
\begin{equation}
w_i = \dfrac{\widetilde{\pi}(X_i)}{q(X_i)}\,.
\end{equation}
Based on $Z$ being known or unknown, weighted averages of $h(X_i)$ yield estimators of $\mu_h$, which we generically denote as $\mu^*_h$. Often, $\mu_h^*$ is the unnormalized IS (UIS) estimator or the self-normalized IS (SNIS) estimator. Other IS estimators have also been proposed  \citep[see][]{vehtari2015pareto, elvira2019generalized,kuntz2021product, martino2018comparison}.

In addition to the target and proposal distributions, variability in $\mu^*_h$ is  critically dependent on: (i) the function of interest $h$ and (ii) the choice of the estimator. Almost all methods assessing the quality of an IS algorithm only utilize the importance weights and do not factor $h$ or the estimator employed to estimate $\mu_h$. A key question that thus remains unanswered is, how should $n$ be chosen so that $\mu^*_h$ is a \emph{good} estimate of  $\mu_h$?

Common diagnostics for IS algorithms measure the discrepancy between the target and proposal distributions.  \cite{chatterjee2018sample} obtain  sample size requirements by utilizing the Kullback-Leibler divergence between $\pi$ and $q$. \cite{sanz2018importance} and \cite{sanz2021bayesian} leverage the $\chi^2$ divergence between $\pi$ and $q$ to quantify the variance of the weights and obtain an estimate of the necessary sample size.
These methods are useful in understanding the quality of the weights, but since they are  invariant to the choice of $h$ and $\mu_h^*$, they are not equipped to directly explain the quality of estimation of $\mu_h$ by $\mu_h^*$. 

Another popular diagnostic in IS is the effective sample size (ESS) which is meant to provide the number of iid samples from $\pi$ that would yield the same variability in $\bar{\mu}_h$ as the variability in $\mu_h^*$.  Using a series of simplifying assumptions, \cite{kong1992note} provides a popularly employed estimator of ESS.  As described by \cite{elvira2018rethinking}, the simplifying assumptions make it so that the resulting estimator does not satisfy key desirable properties. As a consequence, although a reasonable diagnostic for assessing the suitability of the proposal distribution, the estimator fails to truly assess the quality of estimation of $\mu_h$. Nevertheless, it continues to be used in IS as a practical diagnostic owing to reasonable statistical properties, ease of implementation, and lack of better alternatives  \citep[see][for more details]{elvira2018rethinking}.
 
Due to interest in several expectations, often $\mu_h$ and its estimator are multivariate. Naturally, the correlation among components of $\mu^*_h$ impacts the quality of estimation. Since most IS diagnostics do not depend on the choice of the function $h$, this correlation is left ignored.

The contributions of this paper are twofold. First, we present a multivariate analogue of the original definition of ESS, avoiding most of the simplifying assumptions of \cite{kong1992note}. The proposed metric, that we call M-ESS, acknowledges that changing  $q$, $h$, and the estimator employed ${\mu}^*_h$ should yield different quality of estimation. Second, we adapt and integrate multivariate sequential stopping rule techniques of \cite{glynn:whitt:1992} to determine when enough weighted samples have been obtained in IS. Our proposed method stops simulation when the volume of the confidence region for $\mu^*_h$ is small, relative to the variability of $h$ under $\pi$. We show that the confidence region constructed at the random time of termination is asymptotically valid, i.e.,  the probability that $\mu_h$ is contained in the $100(1-\alpha)\%$ confidence region at termination is asymptotically $1-\alpha$. Moreover, adapting ideas from the MCMC literature \citep{vats2017multivariate}, we show that this stopping rule is asymptotically equivalent to stopping the IS simulation when the estimated M-ESS is larger than an \emph{a priori} obtained lower bound. The proposed methodology provides users a clear guideline on when to stop their simulations while redeeming many shortcomings of the current practices.

We implement our proposed ESS and stopping rules in {three} examples. First, we set both $\pi$ and $q$ to be  multivariate Gaussian densities and analytically obtain the true variance of two different IS estimators. Using these variances, we analyze the quality of estimation of M-ESS and the performance of our termination rule as a function of the problem dimension and the degree of correlation in the resulting estimators. Our second example is of a Bayesian Weibull multi-step step-stress model. Here, we employ our ESS for two different multivariate $h$ of interest to demonstrate the difference in the quality of estimation for different choices of $h$. {Our third example is a relatively large-dimensional approximate Bayesian inference problem using IS and integrated nested Laplace approximation, where the quality of estimation of our proposed M-ESS is upheld, despite the problem complexity.}

\section{Importance Sampling and Effective Sample Size}

We recall that interest is in estimating $\mu_h =  \text{E}_{\pi} \left[h(X)\right] \in \mathbb{R}^p$ where $\pi$ may be such that 
\begin{equation}
    \widetilde{\pi}(x) = Z \pi(x)\,.
\end{equation}
We assume that $\widetilde{\pi}$ is known (can be evaluated at any $x \in \mathcal{X}$) while $Z$ may be known or unknown; without loss of generality, we assume $Z = 1$ when $\pi$ can be completely evaluated. For instance, in Bayesian inference, $Z$ is the intractable marginal likelihood. Based on the availability of $Z$, one of the two popular IS estimators may be employed. There certainly are situations where $\widetilde{\pi}$ may also be unavailable \citep[see][]{park:haran:2018}; we exclude these from consideration in this work.

Let $q$ be an appropriately chosen proposal distribution and $X_1, \dots, X_n$ be iid samples from $q$. If $Z$ is known, $\mu_h$ may be estimated using the UIS estimator
\begin{equation} \label{eq:UIS}
\widehat{\mu}_h:= \dfrac{1}{n} \sum_{i=1}^{n} h(X_i)w_i\,.
\end{equation}
When $Z$ is unknown, the UIS estimator cannot be calculated, in which case $\mu_h$ can be estimated by the SNIS estimator
\begin{equation} \label{eq:SNIS}
    \widetilde{\mu}_h :=  \dfrac{\sum_{i=1}^{n} h(X_i)w_i}{\sum_{i=1}^{n}w_i}\,.
\end{equation}
If $q(x) > 0$ for all $x$ such that $\pi(x) > 0$, both $\widehat{\mu}_h$ and  $\widetilde{\mu}_h$ converge to $\mu_h$ with probability 1, as $n \to \infty$; see \cite{Owen:2013,robert2013monte} for more details. Recall, we denote any IS estimator of $\mu_h$ as $\mu^*_h$, and will use the specific notations when referring to either UIS or SNIS. Further, often $\mu^*_h$ exhibits asymptotic normality,  so that as $n \to \infty$,
\begin{equation}
  \sqrt{n}(\mu^*_h - \mu_h) \xrightarrow{d} {\mathcal{N}}(0, \Omega^*)\,, \label{eq:clt}
\end{equation}
where $\Omega^* := \lim_{n\to \infty} n \text{Var}_q(\mu^*_h)$.
By a central limit theorem (CLT), asymptotic normality for $\widehat{\mu}_h$ is straightforward, and $\Omega^*$ is denoted by $\Omega_U$.  For the SNIS estimator, when $\Omega^*$ is finite, asymptotic normality holds with  
\begin{equation} \label{eq:SNIS_variance}
 \Omega^*=\Omega :=  \dfrac{\mathbb{E}_q \left[w(X)^2(h(X) - \mu_h)(h(X) - \mu_h)^T \right]}{\mathbb{E}_q \left[w(X) \right]^2}\,.
\end{equation}
The form of $\Omega$ has been presented before in  \cite{nilakanta2020output,Owen:2013}, but for the sake of completion, we provide the details in the Appendix. Under a finite second moment condition with $\Var_{\pi}(h(X))$ denoted by $\Sigma$, the vanilla Monte Carlo estimator, $\bar{\mu}_h$, also satisfies a CLT  with a limiting covariance matrix
\begin{equation}
\lim_{n\to \infty} n \text{Var}_{\pi} (\bar{\mu}_h) =: \Sigma \,.
\end{equation}

For a univariate $h$, in order to quantify the relative quality of $\mu^*_h$ as compared to $\bar{\mu}_h$, \cite{kong1992note} attempts to define the ESS as the ratio of the respective {mean-squared errors}. Since the bias of the SNIS estimator is difficult to ascertain, ESS is first described as $n \cdot {\Var(\bar{\mu}_h)}/{\Var(\mu^*_h)}$ (in the univariate case, both variances are scalars). Through a series of approximations, \cite{kong1992note} obtains the following estimator of the above ESS:
\begin{equation}\label{eq:kong_ess}
    K_n := \dfrac{1}{\sum_{i=1}^n \bar{w}_i^2}\,,
\end{equation}
where $\bar{w}_i = w_i /\sum_{i=1}^{n} w_i$ are the normalized weights. \cite{elvira2018rethinking} provide a full derivation of \eqref{eq:kong_ess}, including the assumptions and approximations, and discuss various shortcomings of $K_n$. A critical undesirable quality of $K_n$ is that it does not depend on the function of interest $h$ and thus claims the same estimation quality irrespective of $h$. Consequently, $K_n$ is also unable to acknowledge multivariate estimation in $h$. This may be an advantage in some scenarios, e.g., when there is not a particular targeted integral $\mu_h$ and IS is used to build a weighted particle approximation of the target. However, IS is mostly used as a variance reduction technique for the approximation of integrals, and thus, $K_n$ not depending on $h$ is a  limitation in most applications of interest. For instance, in machine learning, IS is often used to estimate the gradient, i.e., a multivariate $h$,  \citep{mohamed2020monte}. Moreover, in rare events events estimation, $h$ can be a multivariate function composed of indicator functions \citep{owen2019importance,miller2021rare}, and it is crucial to capture the suitability of the samples to the function of interest.   Finally, since $1 \leq K_n \leq n$, the estimator is unable to detect improvements in statistical efficiency, a key feature of IS.

We propose a multivariate extension of the definition of ESS based on the original intention of \cite{kong1992note}, but replace the finite-time variances with the estimable limiting variances. That is, we define the  multivariate ESS as
\begin{equation} \label{eq:pop_ESS}
    \text{M-ESS} :=  n\dfrac{\abs{\Sigma}^{1/p}}{\abs{\Omega^*}^{1/p}} \,.
\end{equation}
The $p$th root of determinant of covariance matrices is a dimension-free measure of variability \citep{sengupta1987tests} and is natural since the generalized variability of a random vector is quantified  by the determinant of the corresponding covariance matrix \citep{wilks1932certain}.  {\cite{vats2017multivariate} employ a similar metric in the context of MCMC where $\Omega^*$ is the asymptotic variance in the Markov chain CLT of MCMC ergodic averages. This unified formulation of ESS allows us to compare the efficiency of different sampling methods whenever a meaningful estimator of $\Omega^*$ can be constructed. This is discussed in detail in Section~\ref{ex:lasso}.}

{We present estimators of $\Sigma$ and $\Omega^* = \Omega$ for SNIS.} Since
\begin{equation*}
    {{\Sigma}} := \mathbb{E}_{{\pi}}\left[ (h(X) - \mu_h) (h(X) - \mu_h)^T \right] = \mathbb{E}_q \left[w(X)(h(X) - \mu_h)(h(X) - \mu_h)^T \right]\,,
\end{equation*}
a plug-in estimator of $\Sigma$ is
\begin{equation} \label{eq:var_target}
    \widehat{\Sigma} :=  \sum_{i=1}^{n}\bar{w}_i (h(X_i) - \widetilde{\mu}_h)(h(X_i) - \widetilde{\mu}_h)^T.
\end{equation}

Similarly, for $\Omega$ from \eqref{eq:SNIS_variance}, a multivariate extension of the plug-in estimator by \cite{Owen:2013} is
 \begin{equation}\label{eq:empirical_SNIS_variance}
     \widehat{\Omega}= \dfrac{n^{-1}\sum_{i=1}^{n}w_i^2 (h(X_i) - \widetilde{\mu}_h)(h(X_i) - \widetilde{\mu}_h)^T}{\left(n^{-1}\sum_{i=1}^{n}w_i\right)^2} = n\sum_{i=1}^{n}\bar{w}_i^2 (h(X_i) - \widetilde{\mu}_h)(h(X_i) - \widetilde{\mu}_h)^T\,.
 \end{equation}

\begin{remark}
For UIS, estimators for $\Sigma$ and $\Omega_U$, denoted by $\widehat{\Sigma}_U$ and $\widehat{\Omega}_U$, respectively, are:
\begin{equation*}
    \widehat{\Sigma}_U := \dfrac{1}{n} \sum_{i=1}^{n}w_i(h(X_i) - \widehat{\mu}_h)(h(X_i) - \widehat{\mu}_h)^T \,\,\, \text{ and } \,\,\,
    \widehat{\Omega}_U := \sum_{i=1}^{n}(w_i h(X_i) - \widehat{\mu}_h)(w_i h(X_i) - \widehat{\mu}_h)^T.
\end{equation*}
\end{remark}

{Using the estimators described above, a natural plug-in estimator of the M-ESS is
\begin{equation}
    \label{eq:mess-est}
    \widehat{\text{M-ESS}} = n \dfrac{|\hat{\Sigma}|^{1/p}}{ |\hat{\Omega}_*|^{1/p}}\,.
\end{equation} 
\begin{remark}
The quality of estimation of both $K_n$ and $\text{M-ESS}$ depends critically on the  weights. Since $\text{M-ESS}$ requires estimation of second moments, we  recommend that users choose their ideal proposal before implementing our stopping rules to determine when to stop their simulation.
\end{remark}
\begin{remark}
Throughout the manuscript we assume that IS and SNIS estimators exhibit asymptotic normality. This may not always be true, in which case $\Omega^*$ will be infinite. If $\Sigma$ is finite, then after some $n^*$ initial samples, $\widehat{\text{M-ESS}}/n$ should drop down to 0. We comment more on $n^*$ in the next section.
\end{remark}
}
 
The proposed $\widehat{\text{M-ESS}}$ indicates the quality of estimation of $\mu^*_h$ relative to the vanilla Monte Carlo estimator. Moreover, as we discuss in next section, it also serves as basis the for a principled stopping  criterion and can greatly alleviate ambiguity in terminating simulation.

\section{Using ESS to Stop Simulation} \label{sec:ssr}

A key practical question in any simulation paradigm is, when should sampling stop? Sequential stopping rules that check whether a desired criteria has been satisfied, have been useful in answering this question in steady-state simulations \citep{dong2019new,glynn:whitt:1992}, stochastic programming \citep{bayraksan2012fixed}, general Monte Carlo \citep{frey2010fixed, vats:2021}, and MCMC \citep{flegal2015relative,vats2017multivariate}. Such stopping rules terminate simulation when the size of the confidence region of the estimator is small. {For the rest of the paper, we focus our attention on the SNIS estimator (unless stated otherwise).}

Ellipsoidal large-sample confidence regions for $\mu_h$ are available due to \eqref{eq:clt} and \eqref{eq:empirical_SNIS_variance}. Let $\chi^2_{1-\alpha,p}$ denote the $(1-\alpha)$-quantile for a chi-squared distribution with $p$ degrees of freedom. Then a large-sample $100(1-\alpha)\%$ confidence region around $\widetilde{\mu}_h$ is,
\begin{equation}
\label{eq:confidence_region}
C_{\alpha}(n) = \left\{\mu_h \in \mathbb{R}^p: n(\widetilde{\mu}_h - \mu_h)^T \widehat{\Omega}^{-1}(\widetilde{\mu}_h - \mu_h) < \chi^2_{1-\alpha, p} \right\}\,.
\end{equation}
The volume of the confidence region is,
\begin{equation}
\label{eq:vol}
   \text{Vol}({C_{\alpha}(n)}) = \dfrac{2 \pi^{p/2}}{p \Gamma(p/2)} \left(\dfrac{\chi^2_{1-\alpha, p}}{n}\right)^{p/2} |\widehat{\Omega}|^{1/2}\,. 
\end{equation}
For a user-chosen $\epsilon$-tolerance, we derive a stopping rule that terminates the simulation when the estimated ESS is larger than a pre-determined lower-bound. {In other words,} consider stopping the simulation when the $p$th root of the volume of $C_{\alpha}(n)$ is an $\epsilon$th fraction of $|\widehat{\Sigma}|^{1/2p}$. Specifically, for $s(n) \to 0$ as $n\to \infty$, the process terminates the first time when
\begin{equation}
\label{eq:stop_theory}
\text{Vol}({C_{\alpha}(n)})^{1/p}  +  s(n) \leq \epsilon |\widehat{\Sigma}|^{1/2p}\,.
\end{equation}
Here, $s(n)$ is chosen to ensure that a user-chosen minimum simulation effort of $n^*$ is guaranteed so that the simulation does not end prematurely due to unstable early estimates. We will comment on a particular choice later.  {Equation~\eqref{eq:stop_theory} is the relative standard deviation sequential stopping rule of \cite{vats2017multivariate}.} Ignoring $s(n)$ for large $n$, \eqref{eq:vol} implies that the stopping rule in  \eqref{eq:stop_theory} is equivalent to stopping the simulation when:
\begin{equation}
   \widehat{\text{M-ESS}} \geq \dfrac{2^{2/p} \pi}{(p \Gamma(p/2))^{2/p}} \dfrac{1}{\epsilon^2}\, \chi^2_{1-\alpha, p}=: L_{\alpha, \epsilon,p}\,. \label{eq:stop_practical}
\end{equation}
This reformulation furnishes a rule that terminates the IS process when  $\widehat{\text{M-ESS}}$ is larger than $L_{\alpha,\epsilon,p}$. A lower level of relative tolerance, $\epsilon$, yields a higher $L_{\alpha,\epsilon,p}$.   Notice that the lower bound $L_{\alpha,\epsilon,p}$ can be calculated even before the simulation begins. Implementations that yield large variability in the SNIS estimator will require more sample size to reach the desired lower bound.  Such sequential stopping rules terminate simulation at a random time, and thus additional care must be taken to ensure asymptotic validity of the resulting confidence regions is retained. The following theorem is built on the works of \cite{glynn:whitt:1992,vats2017multivariate} to establish this asymptotic validity in the context of IS. The theorem and proof (in the Appendix) is presented for the SNIS estimator; an analogous statement and proof is available for the UIS estimator.

\begin{theorem} \label{th:ssr}
Let $R_n(X)$ be a strongly consistent estimator of an attribute of the system, $R(X)$. That is, let $R_n(X) \to R(X)$ with probability 1 as $n \to \infty$. Define $s(n) = \epsilon R_n(X)I(n < n^*) + n^{-1}$ for some finite $n^* > 0$. Additionally, note that $\widehat{\Omega} \to \Omega$ as $n\to \infty$ with probability 1. For $\epsilon > 0$, consider the stopping rule
\begin{equation} \label{eq:stopping_rule}
    T^*(\epsilon) = \inf \left\{n \geq 0: \text{Vol}({C_{\alpha}(n)})^{1/p} + s(n) \leq \epsilon R_n(X) \right\}\,.
\end{equation}
As $\epsilon \to 0$, $T^*(\epsilon) \to \infty$ and $\Pr\{\mu_h \in C_{\alpha}(T^*(\epsilon))\} \to 1-\alpha$.
\end{theorem}
Due to the random-time termination, \cite{glynn:whitt:1992} show that strong consistency of $\widehat{\Omega}$ and $R_n(X)$ is necessary for asymptotic validity of the resulting confidence regions. When $n < n^*$, $s(n) = \epsilon R_n(X) + n^{-1}$, making it so that simulation cannot stop. Thus, this choice of $s(n)$ ensures termination occurs after a minimum simulation effort of $n^*$. Setting $R(X) = |\Sigma|^{1/2p}$ and  $R_n(X)  = |\widehat{\Sigma}|^{1/2p}$, $T^*(\epsilon)$ is equal to the smallest $n$ that satisfies \eqref{eq:stop_theory}. Thus, stopping the simulation by checking whether $\widehat{\text{M-ESS}} > L_{\alpha, \epsilon, p}$ will produce adequate confidence regions. The relative precision, $\epsilon$, may be chosen depending on the desired quality of estimation. As $\epsilon \to 0$, $L_{\alpha, \epsilon, p} \to \infty$, implying the required sample size would also increase to infinity.

\section{Examples} \label{sec:examples}

We implement our proposed multivariate ESS in three different examples\footnote{Reproducible code for the examples is available at https://github.com/medhaaga/Importance-Sampling-Stopping-Rule}. First, we present a controlled scenario where the groundtruth is available, and therefore validation of the proposed methodology is possible. We estimate the mean vector of a multivariate normal distribution with a multivariate normal proposal using both UIS and SNIS. We arrive at an expression for the true $\Omega^*$ and use this to assess the proposed termination criterion. Next, we present a Bayesian multi-step step-stress model where the interest is in estimating two different functions $h$. We demonstrate the utility of our $\widehat{\text{M-ESS}}$ in acknowledging these different estimation goals. {Our last example is that of approximate Bayesian inference using IS and integrated nested Laplace approximation (INLA). We highlight that in this relatively large-dimensional problem, our proposed M-ESS is well-behaved and particularly well-suited for comparisons with MCMC.}

\subsection{Multivariate Normal} \label{ex:normal}

Let $\pi = {\mathcal{N}}(\mu, \Lambda)$ where $\mu \in \mathbb{R}^p$ and $\Lambda$ is a $p \times p$ positive-definite matrix and suppose $h$ is the identity function so that the goal is to estimate $\mu_h = \mu$.  Set the proposal to be $q = N(\mu, \Upsilon)$. The limiting variance of the SNIS estimator $\widetilde{\mu}_h$ is available (details are in the Appendix):
\begin{equation} \label{eq:multinorm-SNISvariance}
   \Omega =  \dfrac{|\Upsilon|^{1/2}}{|\Lambda| |2\Lambda^{-1} - \Upsilon^{-1}|^{1/2}} (2\Lambda^{-1} - \Upsilon^{-1})^{-1}\,.
\end{equation}
Additionally, for the UIS estimator 
 \begin{equation} \label{eq:multinorm-UISvariance}
     \Omega_U = \dfrac{|\Upsilon|^{1/2}}{|\Lambda| |2\Lambda^{-1} - \Upsilon^{-1}|^{1/2}}\left[ (2\Lambda^{-1} - \Upsilon^{-1})^{-1} + \mu \mu^T\right] - \mu \mu^T\,.
 \end{equation}

\begin{figure}[htbp]
    \centering
    \includegraphics[width = .3\linewidth]{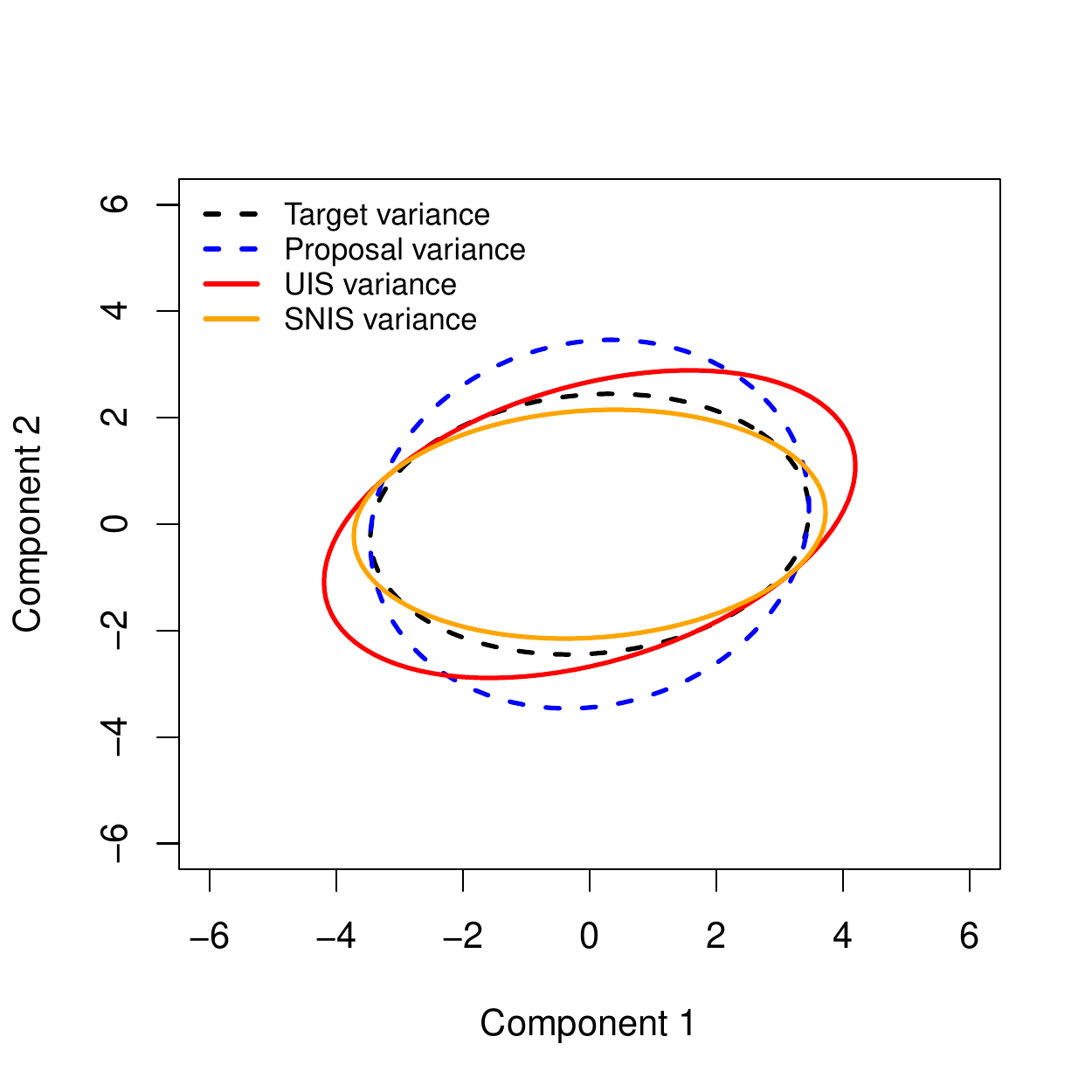}
    \includegraphics[width = .3\linewidth]{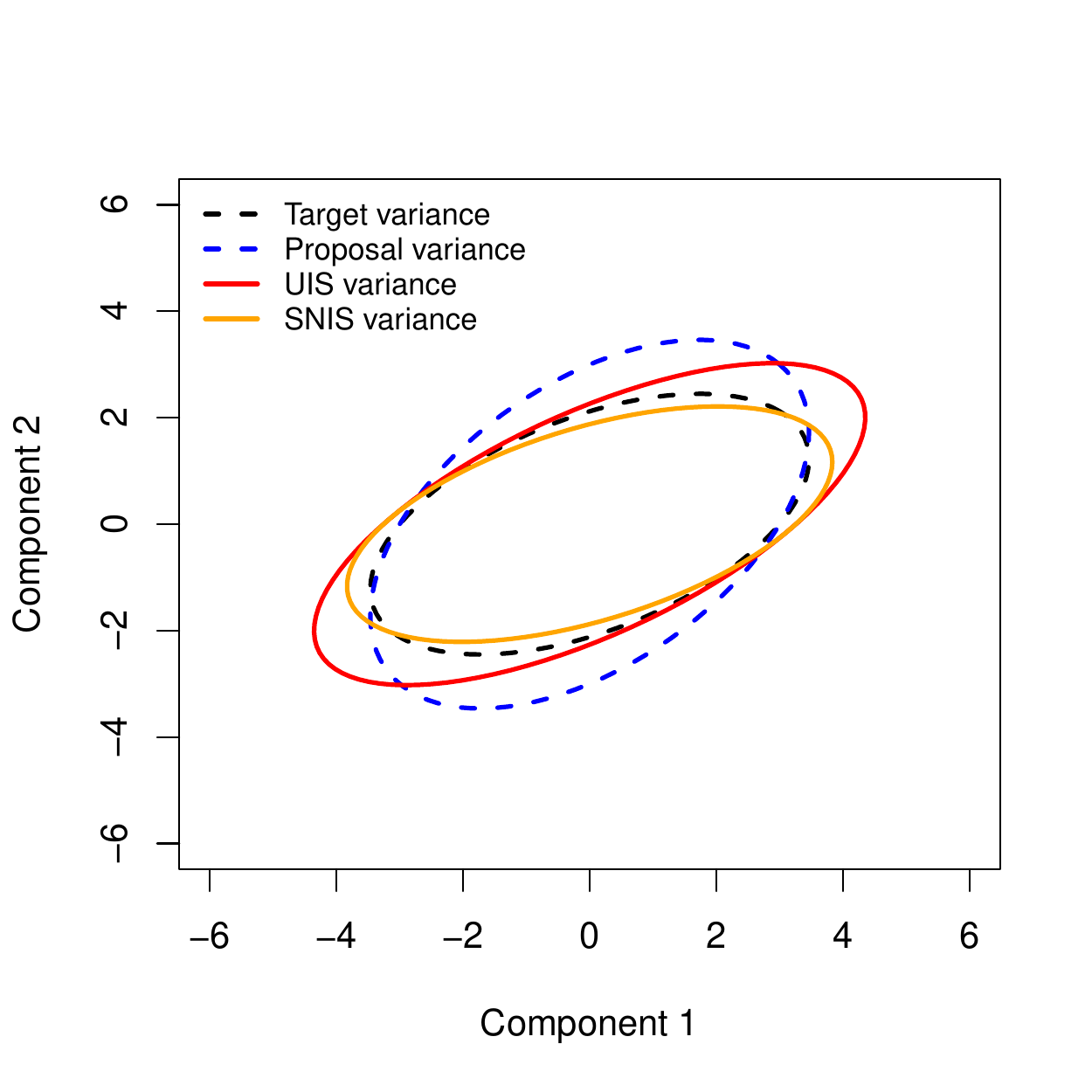}
    \includegraphics[width = .3\linewidth]{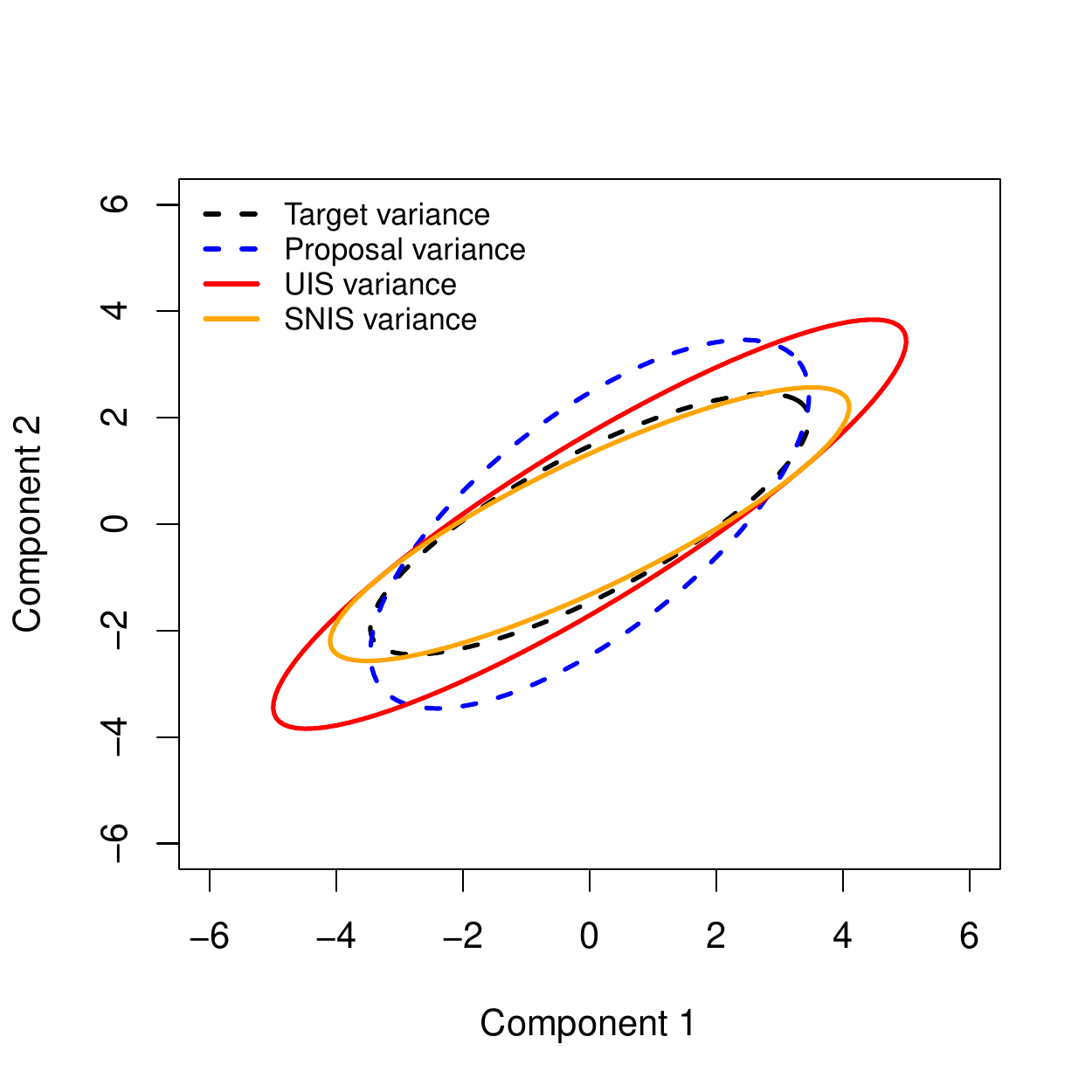}
    
    \includegraphics[width = .3\linewidth]{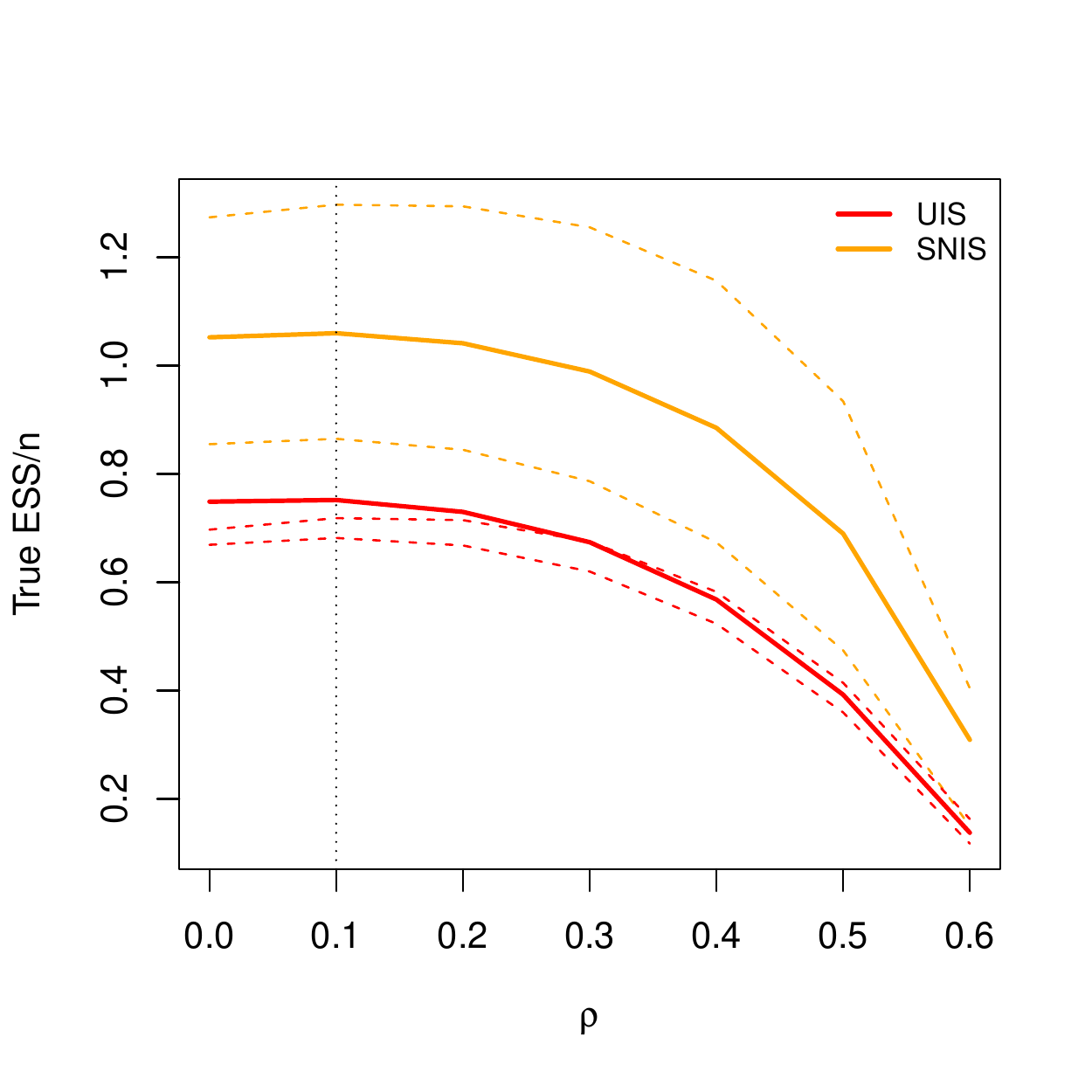}
     \includegraphics[width = .3\linewidth]{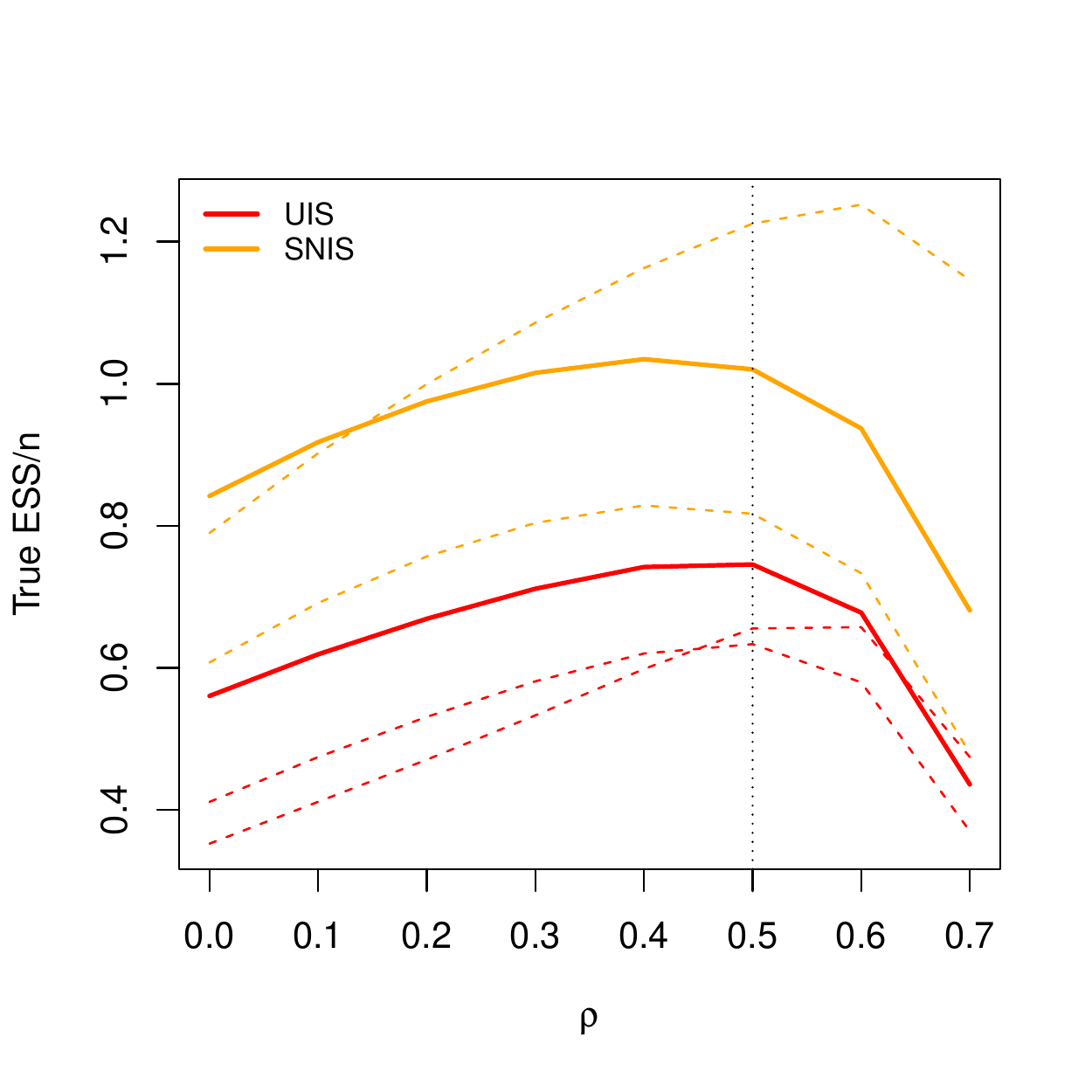}
    \includegraphics[width = .3\linewidth]{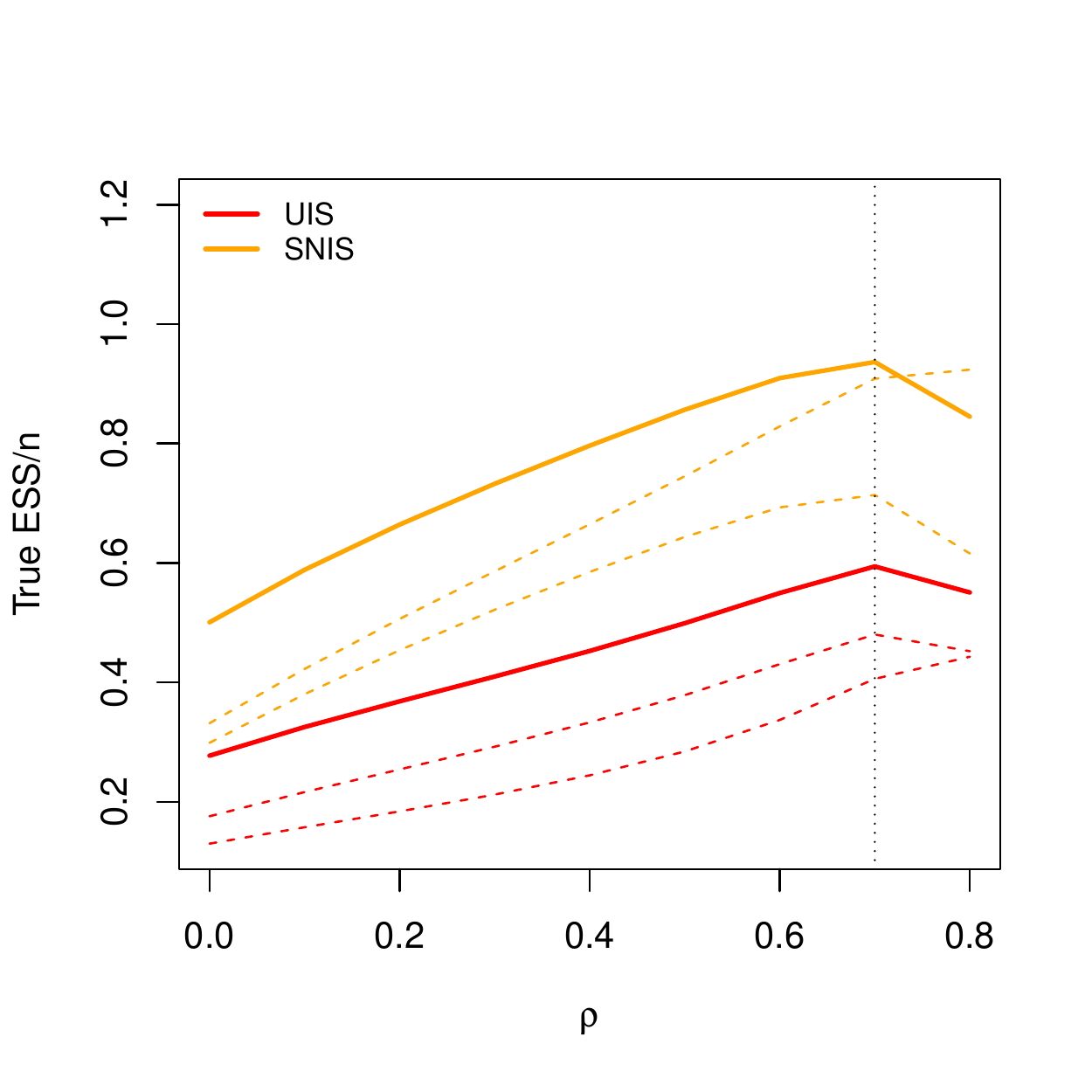}

    \caption{(Top row) The target, proposal, UIS, and SNIS covariance ellipses for Settings 1, 2, 3 (left to right). (Bottom row) True ESS$/n$ for SNIS and UIS from \eqref{eq:multinorm-SNISvariance} and \eqref{eq:multinorm-UISvariance}  for each setting. The vertical dotted line marks the value of $\rho$ chosen in the three settings.}
    \label{fig:biNormal-trueESSvsRho}
\end{figure}

We first fix $p = 2$ and for $\sigma_1, \sigma_2, \sigma > 0$ and $0< \lambda, \rho <1$, we consider $\Lambda$ and $\Upsilon$ of the following form:
\[
\Lambda = \begin{pmatrix}
\sigma_1 & \lambda \sqrt{\sigma_1 \sigma_2}\\
\lambda \sqrt{\sigma_1 \sigma_2} & \sigma_2
\end{pmatrix} \qquad \text{ and } \qquad \Upsilon = \begin{pmatrix}
\sigma & \rho \sigma\\
\rho \sigma & \sigma
\end{pmatrix}\,.
\]
We compare the various ESSs for SNIS and UIS under three different settings:
\begin{enumerate}
    \item \textit{Setting 1}: low target and proposal correlation, with $\lambda = \rho = 0.1$:
    \begin{equation}\label{eq:multinorm-setting1}
        \Lambda = \begin{pmatrix}
    2 & 0.1 \sqrt{2}\\
    0.1 \sqrt{2} & 1
    \end{pmatrix} \qquad \qquad \Upsilon = \begin{pmatrix}
    2 & 0.2\\
    0.2 & 2
    \end{pmatrix}\,.
    \end{equation}
    
     \item \textit{Setting 2}: medium correlation in target and proposal, with $\lambda = \rho = 0.5$:
     \begin{equation} \label{eQ:multinorm-setting2}
         \Lambda = \begin{pmatrix}
    2 & 0.5\sqrt{2}\\
    0.5 \sqrt{2} & 1
    \end{pmatrix} \qquad \qquad \Upsilon = \begin{pmatrix}
    2 & 1\\
    1 & 2
    \end{pmatrix}\,.
     \end{equation}
   
    \item \textit{Setting 3}: high and differing correlation, with $\lambda = 0.8$ and $\rho = 0.7$:
    \begin{equation} \label{eq:multinorm-setting3}
        \Lambda = \begin{pmatrix}
    2 & 0.8 \sqrt{2}\\
    0.8 \sqrt{2} & 1
    \end{pmatrix} \qquad \qquad \Upsilon = \begin{pmatrix}
    2 & 1.4\\
    1.4 & 2
    \end{pmatrix}\,.
    \end{equation}
\end{enumerate}
Throughout, we set $\mu = \mathbf{1}_p$. For $p = 2$, the covariance ellipse of $\Lambda$ and $\Upsilon$ in the three settings, along with ellipses for $\Omega$ and $\Omega_U$  are shown in Figure~\ref{fig:biNormal-trueESSvsRho} (top). It is evident that in all three settings the variance of UIS estimator is larger than that  of the SNIS estimator. Figure~\ref{fig:biNormal-trueESSvsRho} (bottom) shows the univariate  and multivariate true ESS for UIS and SNIS  vs the correlation between the two components in the proposal distribution; the chosen $\rho$ for each setting in indicated through a vertical dashed line. First, the SNIS estimator is clearly preferred here over the UIS estimator; this is unsurprising given the ellipses in the top row. Second, the differing quality of univariate and multivariate ESS estimation is apparent. Not accounting for the complex dependence structures (as presented in the top row), leads to inadequate understanding of the overall estimation quality. 
 
Next we demonstrate the performance of the proposed stopping rule in \eqref{eq:stop_practical}. Controlling the desired quality of estimation through $\epsilon$, we study the impact of i) the problem dimension, $p$, ii) the  target/proposal setting, and iii) the choice of estimator (UIS and SNIS). Since dimensions will increase from $p = 2$, we present a general form of $\Lambda$ and $\Upsilon$, similar to the three previous settings:
\begin{equation}
    \Lambda = \begin{pmatrix}
A & B \\
B & C
\end{pmatrix} \qquad
\Upsilon = \begin{pmatrix}
A^\prime & B^\prime \\
B^\prime & C^\prime
\end{pmatrix}\,,
\end{equation}
where 
 \begin{equation*}
A  = \begin{pmatrix}
\sigma_1 & \hdots & \lambda \sigma_1 \\
\vdots & \ddots & \vdots\\
\lambda\sigma_1 & \hdots & \sigma_1
\end{pmatrix}, B = \begin{pmatrix}
\lambda \sqrt{\sigma_1\sigma_2} & \hdots & \lambda \sqrt{\sigma_1\sigma_2}\\
\vdots & \ddots & \vdots\\
\lambda \sqrt{\sigma_1\sigma_2} & \hdots & \lambda \sqrt{\sigma_1\sigma_2}
\end{pmatrix} \text{, and } C = \begin{pmatrix}
\sigma_2 & \hdots & \lambda \sigma_2 \\
\vdots & \ddots & \vdots\\
\lambda\sigma_2 & \hdots & \sigma_2
\end{pmatrix}
\end{equation*}
 \begin{equation*}
  A^\prime  = \begin{pmatrix}
\upsilon_1 & \hdots & \rho \upsilon_1 \\
\vdots & \ddots & \vdots\\
\rho \upsilon_1 & \hdots & \upsilon_1
\end{pmatrix}, B^\prime = \begin{pmatrix}
\rho \sqrt{\upsilon_1\upsilon_2} & \hdots & \rho \sqrt{\upsilon_1\upsilon_2} \\
\vdots & \ddots & \vdots\\
\rho \sqrt{\upsilon_1\upsilon_2} & \hdots & \rho \sqrt{\upsilon_1\upsilon_2}
\end{pmatrix} \text{, and } C^\prime = \begin{pmatrix}
\upsilon_2 & \hdots & \rho \upsilon_2 \\
\vdots & \ddots & \vdots\\
\rho \upsilon_2 & \hdots & \upsilon_2
\end{pmatrix}\,.    
 \end{equation*}

We set $\sigma_1 = 2, \sigma_2 = 1$ and   $\upsilon_1 = \upsilon_2 = 2$. Similar to before, for the three settings, we let 1) $\lambda =  \rho = 0.1$ and 2) $\lambda =  \rho = 0.5$, and 3) $\lambda = 0.8, \rho = 0.7$. These three settings represent low, medium, and high correlation between components for all dimensions.

\begin{figure}[htbp]
    \centering
    
    \includegraphics[width = .32\linewidth]{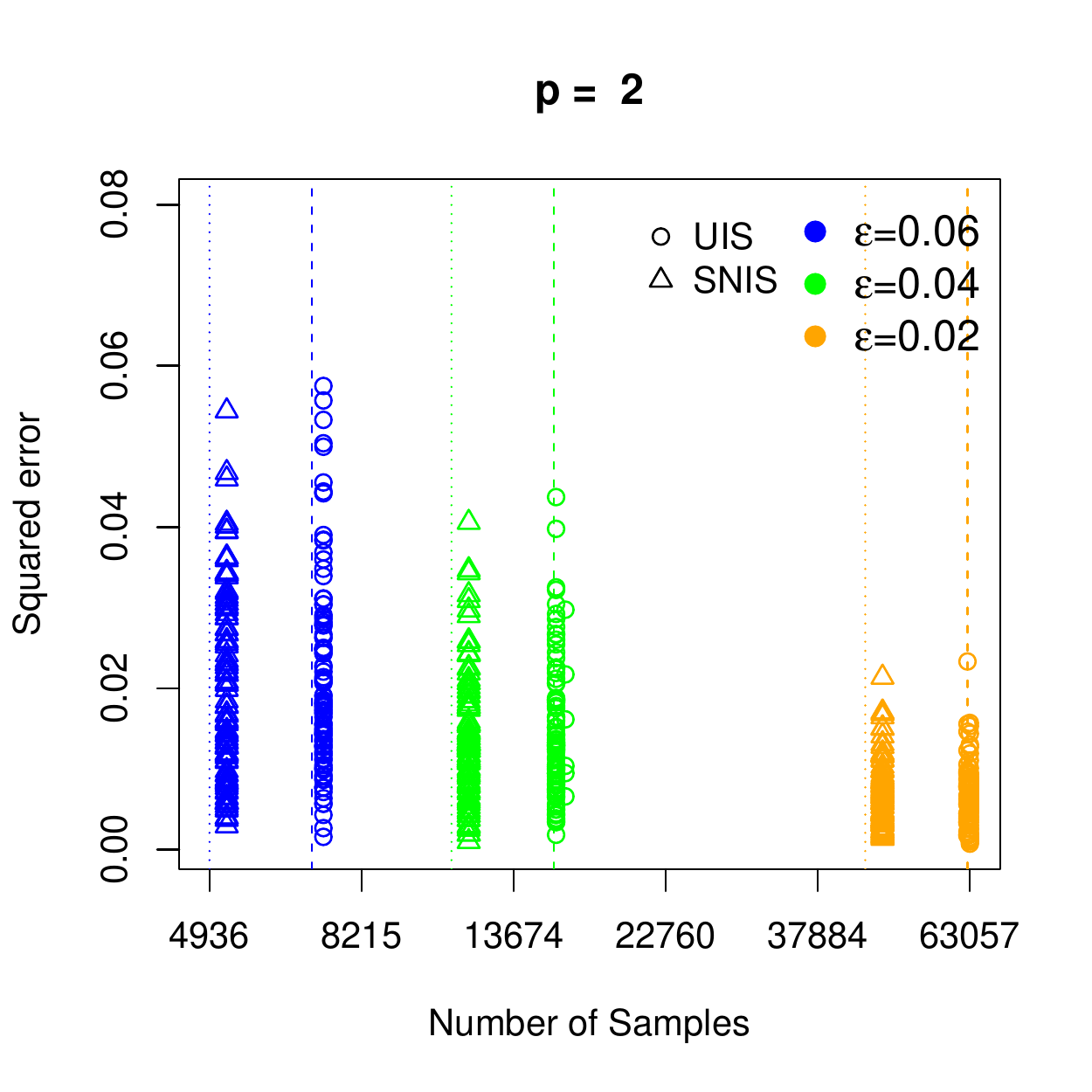}
    \includegraphics[width = .32\linewidth]{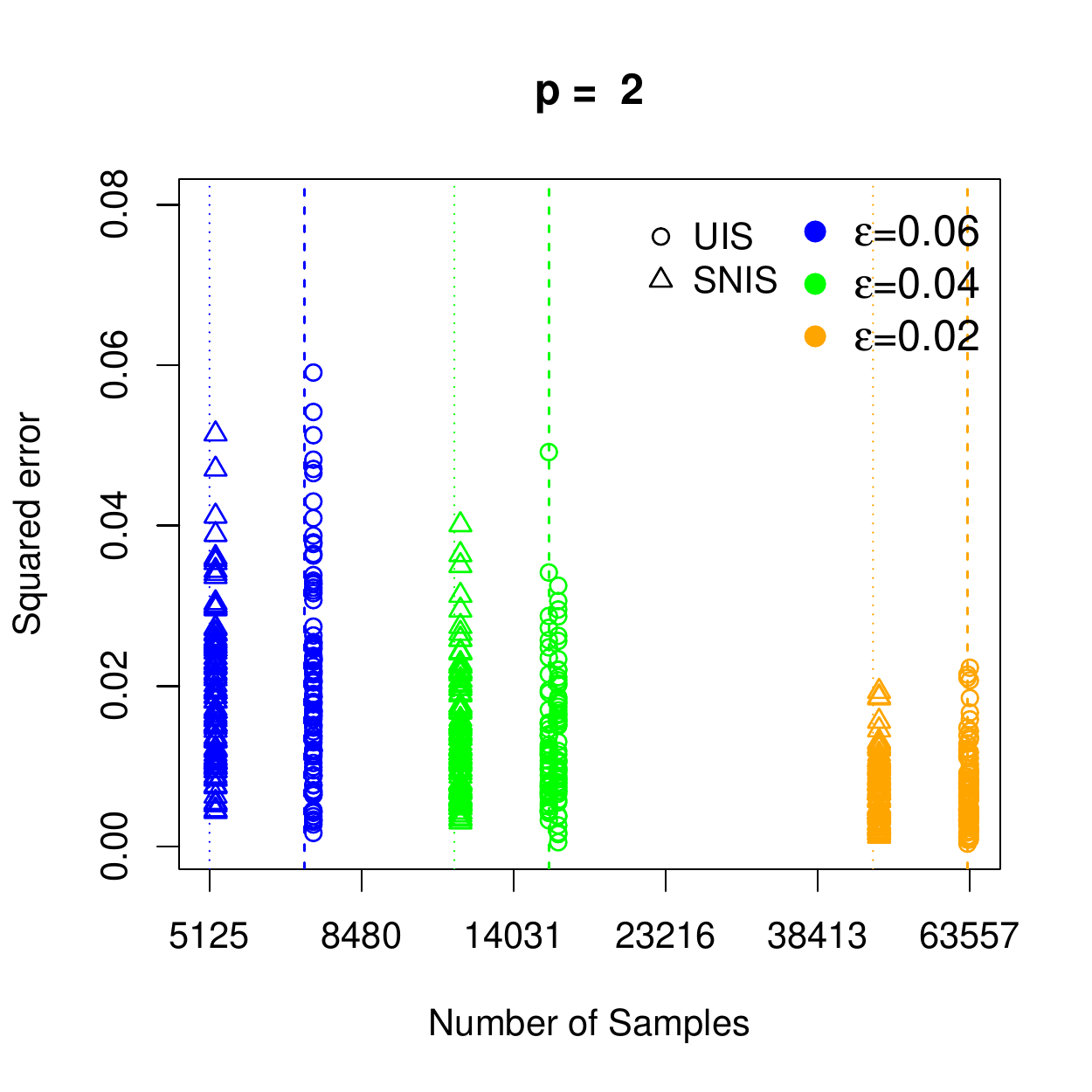}
    \includegraphics[width = .32\linewidth]{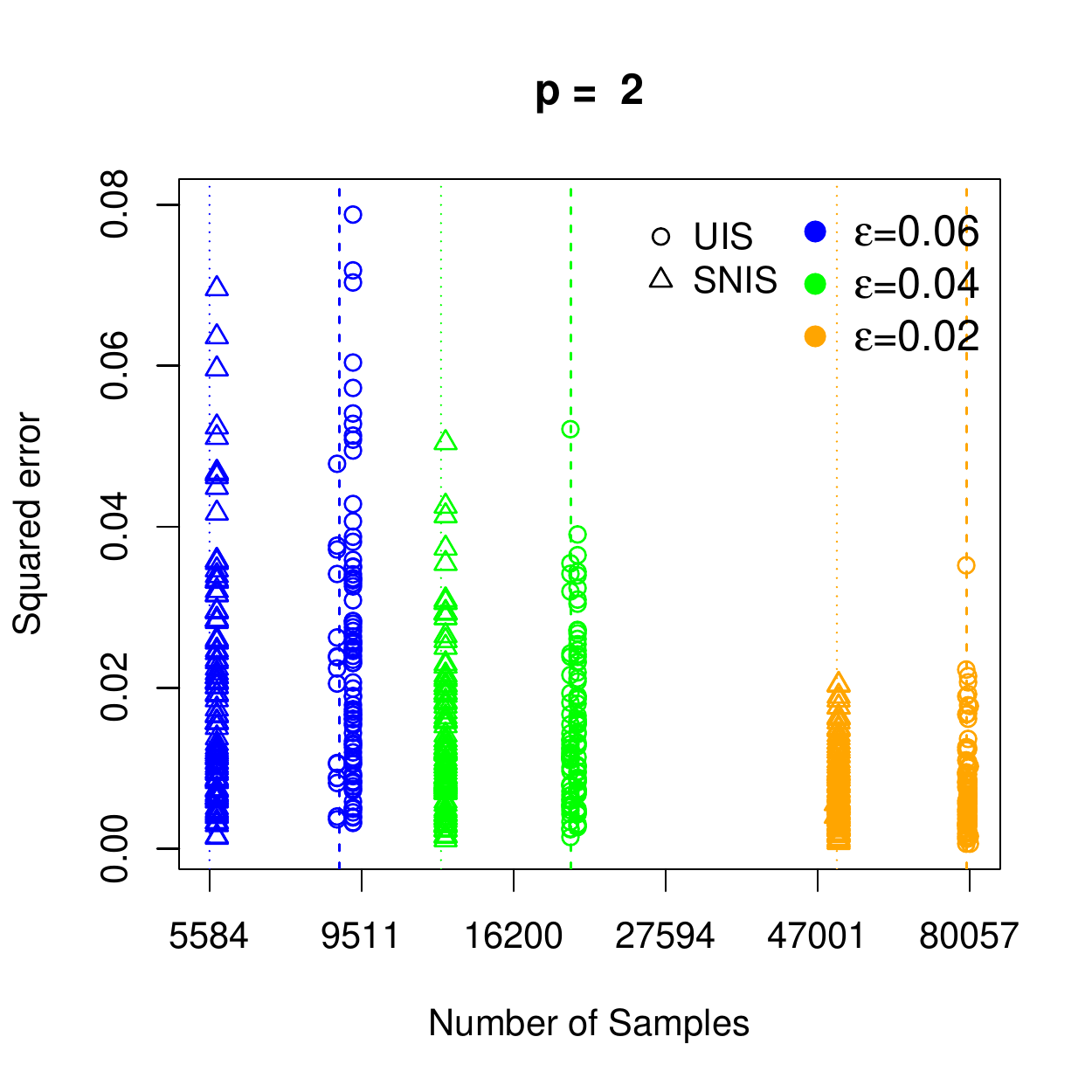}
    \includegraphics[width = .32\linewidth]{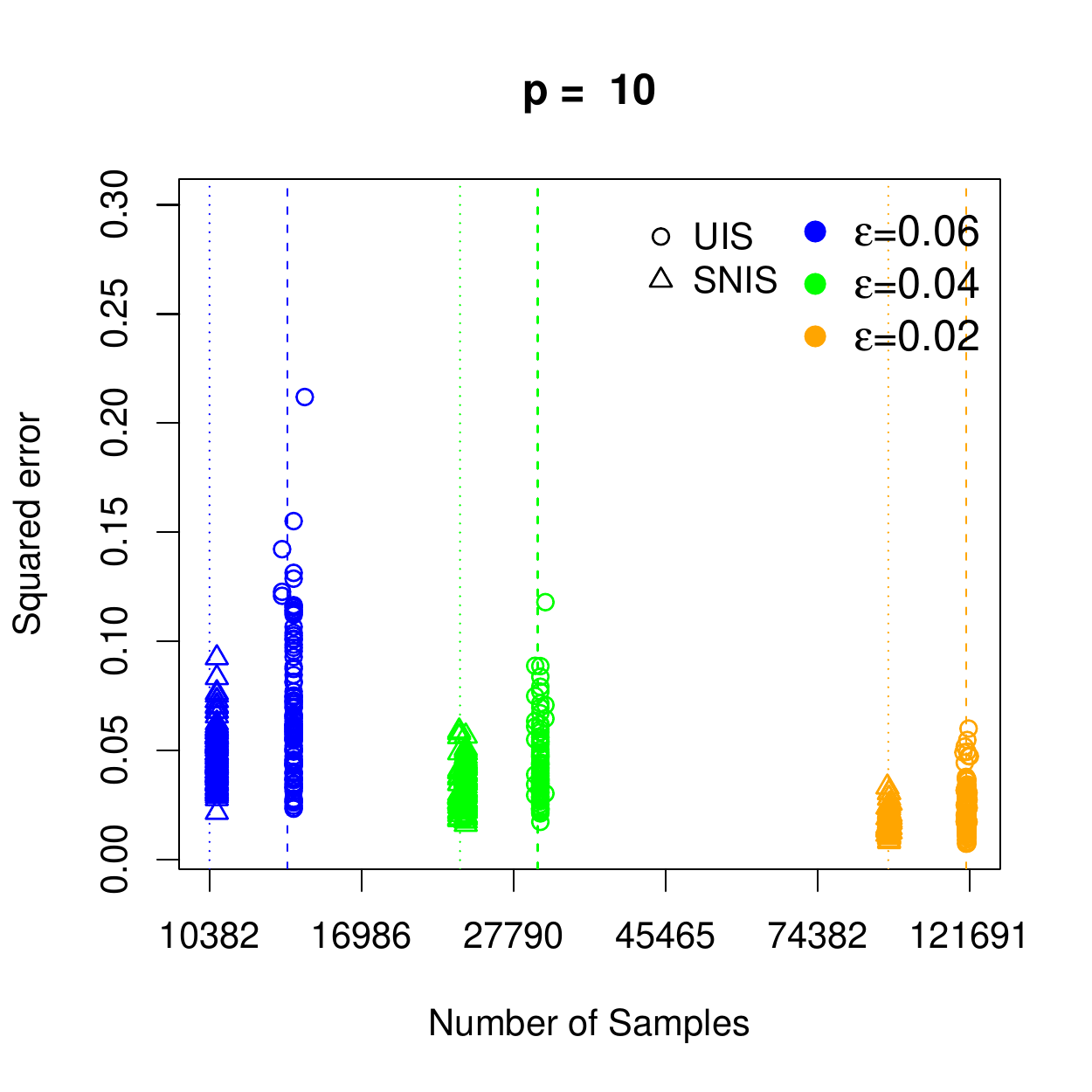}
    \includegraphics[width = .32\linewidth]{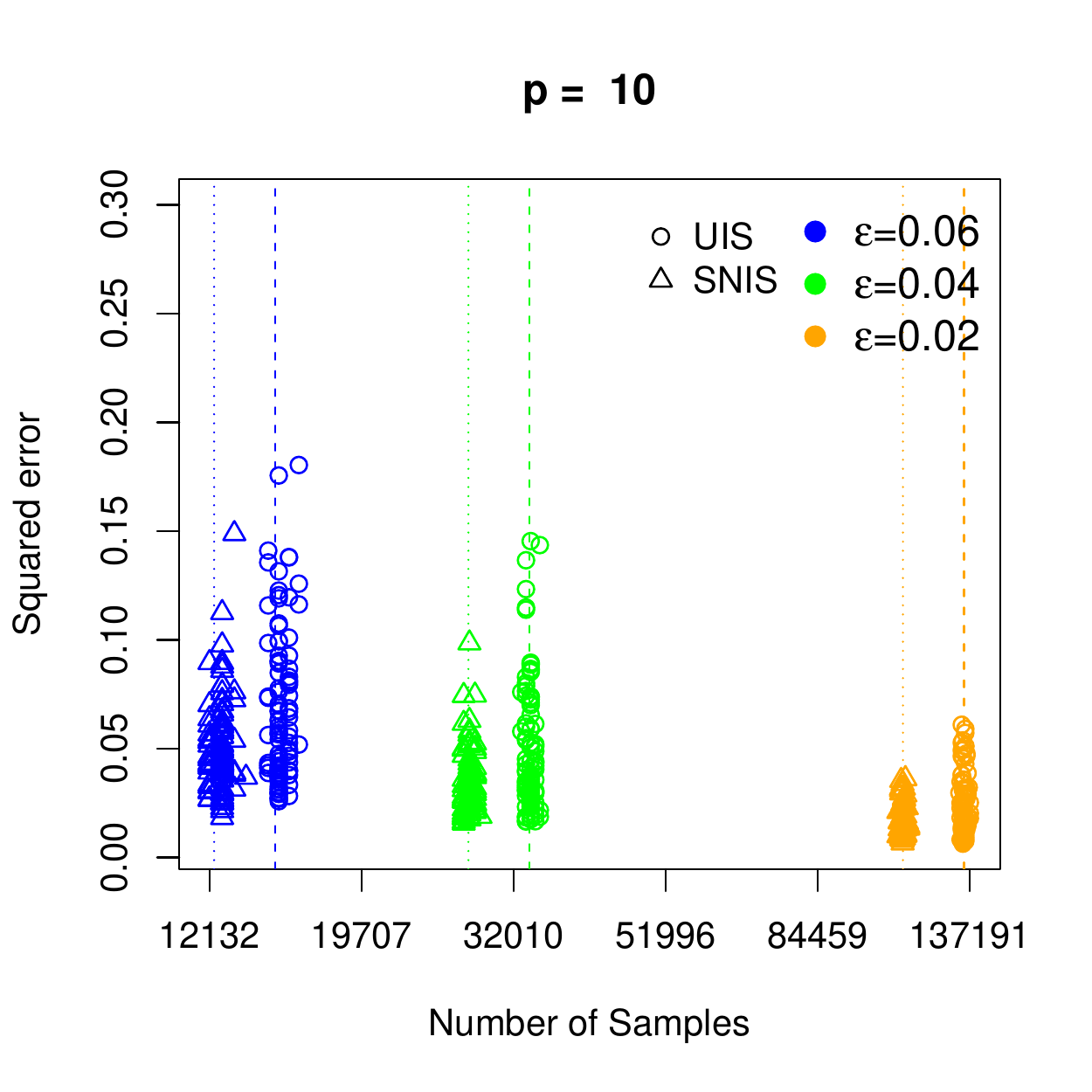}
    \includegraphics[width = .32\linewidth]{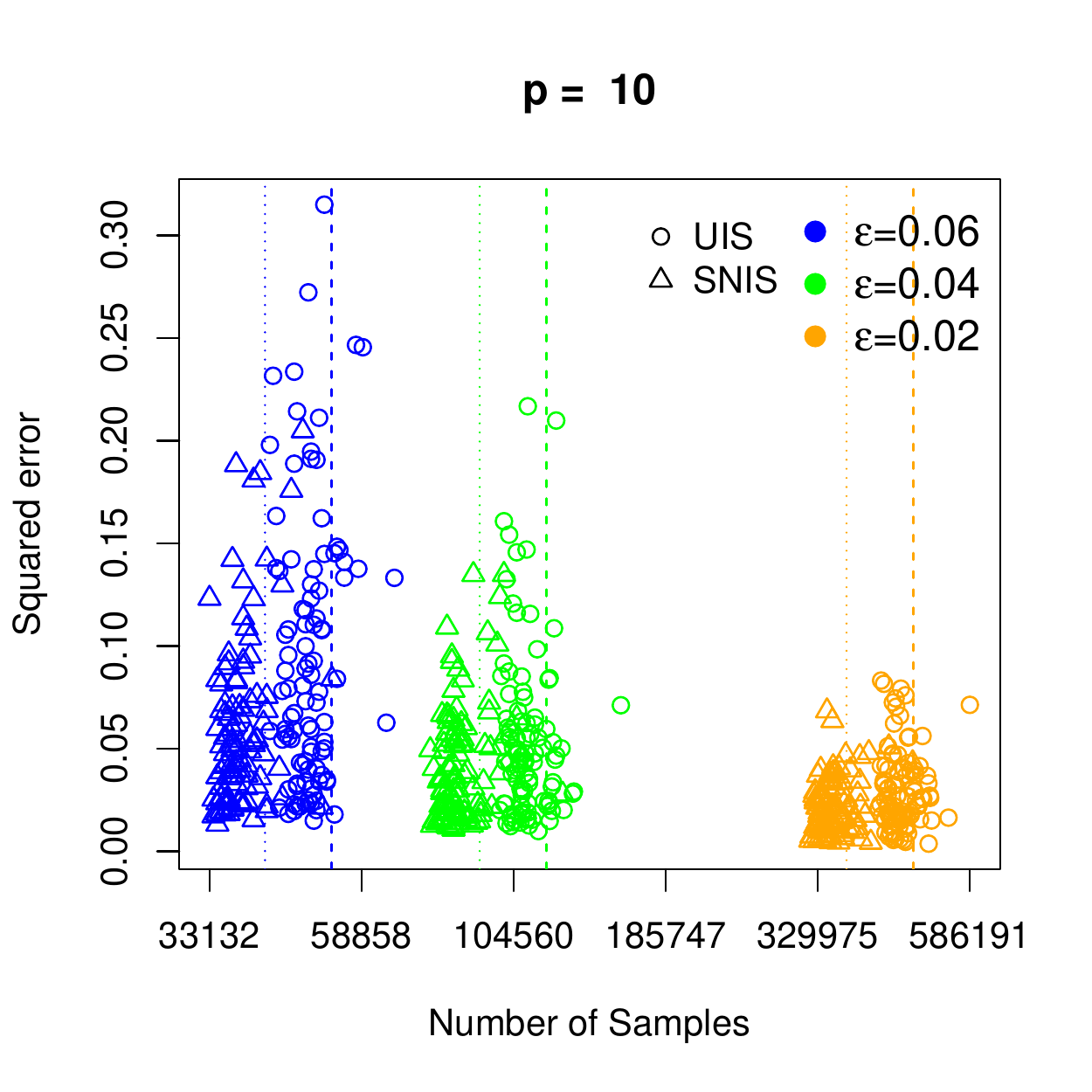}

    \caption{$L_2$ error of SNIS and UIS estimator vs termination time for varying $\epsilon$ for $p = 2 \text{ and } 10$ (top and bottom row).  The truth is marked with  vertical  lines for both estimators.}
    \label{fig:multiNormal-sqeVSiter}
\end{figure}

We implement our multivariate stopping rule in \eqref{eq:stop_practical} in 100 repeated simulations for varying choices of $\epsilon$ for both the UIS and SNIS estimators. Figure~\ref{fig:multiNormal-sqeVSiter} presents the $L_2$ norm of error in estimation of $\mu$ versus the sample size at $\epsilon$-termination. That is, we calculate $\|\mu^* - \mu\|_2$ and plot it versus $n$ at termination for each $\epsilon$. The top row has results for $p = 2$ and the bottom row for $p = 10$. Here the circles represent the IS runs where $\mu^*$ is the UIS estimator and the triangles represent the SNIS estimator. 
First, we note that since UIS produces more inefficient estimators;\textbf{} for any given $\epsilon$, UIS terminates later than SNIS. Additionally, as $\epsilon$ decreases, the variability in the squared error also decreases; this is a direct consequence of an increase in the required sample sizes. Finally, we note that the quality of estimation of the ESS over all three settings is sound. Setting 3, which has the most correlation in the target, exhibits the most variability in the termination time.

\subsection{Weibull Multi-Step Step-Stress Model} \label{ex:weibull}

Consider the fish dataset of \cite{pal2021bayesian}, described in Table~\ref{tab:fish_data}, where the swimming performance of $n = 14$ fish was investigated with an initial swimming rate of 15 cm/sec. The time at which a fish could not maintain its natural position was recorded as the failure time. The flow rate was increased by 5 cm/sec
every time after 110, 130, and 150 minutes. Here, the increased flow rate can be thought of as a stress factor. Thus, there are four stress levels and the observed number of failures at each level, $n_k$, $k = 1,2,3,4$, is $6, 3, 3, \text{ and } 2$, respectively. The failure times are centered by 80 and scaled by 100 as recommended by \cite{pal2021bayesian}, so that time-points corresponding to stress change are  $\tau_1 = 0.3, \tau_2 = 0.5, \tau_3 = 0.7$. 

\begin{table}[htbp]
    \centering
    \begin{tabular}{|c|c|}
\hline
Stress level & Failure times \\
\hline
\hline
    $s_1$ & 83.50, 91.00, 91.00, 97.00, 107.00, 109.50 \\
    \hline
    $s_2$ & 114.00, 115.41, 128.61 \\
    \hline
    $s_3$ & 133.53, 138.58, 140.00\\
    \hline
    $s_4$ & 152.08, 155.10 \\ \hline
\end{tabular}
    \caption{Fish Dataset}
    \label{tab:fish_data}
\end{table}
Let $\bar{n}_j'= \sum_{i=1}^{j}n_i$ be the total number of failures observed during and before the $j$th stress-level and let $t_r$ denote the time-to-failure of the $r$th fish. Let the collected data of ordered observed failure times be denoted by $\mathcal{D}$. \cite{pal2021bayesian} assume that the
lifetime distribution of the experimental units under a given stress level $k$, follows Weibull$(\alpha, \lambda_k)$ for $\alpha, \lambda_k > 0$ with probability density function
\begin{equation}
    f_k(t) = \lambda_k \alpha t^{\alpha - 1} e^{-\lambda_k t^\alpha}\,.
\end{equation}
To allow for ordering in the time-to-failure with subsequent stress levels, the $\lambda$'s are assumed to be ordered so that, $\lambda_1 \leq \lambda_2 \leq \lambda_3 \leq \lambda_{4}$. Let $\theta = (\alpha, \lambda_1, \lambda_2, \lambda_3, \lambda_{4})$. For $a > 0,\, b > 0,\, a_i > 0;\, i = 0, \dots 4$ and $\textbf{a} = (a_1, a_2, a_3, a_4)$, the following are the independent priors assumed by \cite{pal2021bayesian}:
\begin{equation}
    \alpha \sim \text{Gamma}(a,b) \text{ and } \lambda \sim \text{ODG}(a_o, b_o, \textbf{a})\,,
\end{equation}
where ODG stands for the ordered Dirichlet Gamma distribution. Given the Bayesian paradigm, the resulting posterior distribution is the primary object of interest and can be written down as
\begin{align} \label{eq:weibull-posterior}
\pi(\theta | \mathcal{D}) &\propto \pi^\ast_1({\lambda} | \mathcal{D}, \alpha) \pi^\ast_2(\alpha | \mathcal{D}) g(\alpha, \lambda | \mathcal{D})\,,
\end{align}
where $\pi_1^*$ and $\pi_2^*$ are densities of an ODG and Gamma distribution respectively; the exact parameters and details on $g$ are given in the Appendix. We set $a = b = a_o = b_o = 0.5$ and $\textbf{a} = (1, 1, 1, 1)$.

Interest may be in the posterior mean or the average lifetime units for a stress level.  For a Weibull$(\alpha, \lambda)$ distribution, the mean is $\lambda^{-1/\alpha}\Gamma(1 + 1/\alpha)$, so two functions of interest are
\begin{equation} \label{eq:fish-integrands}
    h_1(\theta) = \theta \qquad \qquad h_2(\theta) = \left(\lambda_1^{-1/\alpha} \Gamma(1 + 1/\alpha), \dots , (\lambda_{4}^{-1/\alpha} \Gamma(1 + 1/\alpha) \right)\,.
\end{equation}

Similar to \cite{pal2021bayesian}, we employ IS with proposal
\begin{equation}
   q(\alpha, \lambda) \sim  \pi^\ast_1({\lambda} | \mathcal{D}, \alpha) \pi^\ast_2(\alpha | \mathcal{D})\,. 
\end{equation}
\begin{figure}[htbp]
    \centering
    \includegraphics[width = .26\linewidth]{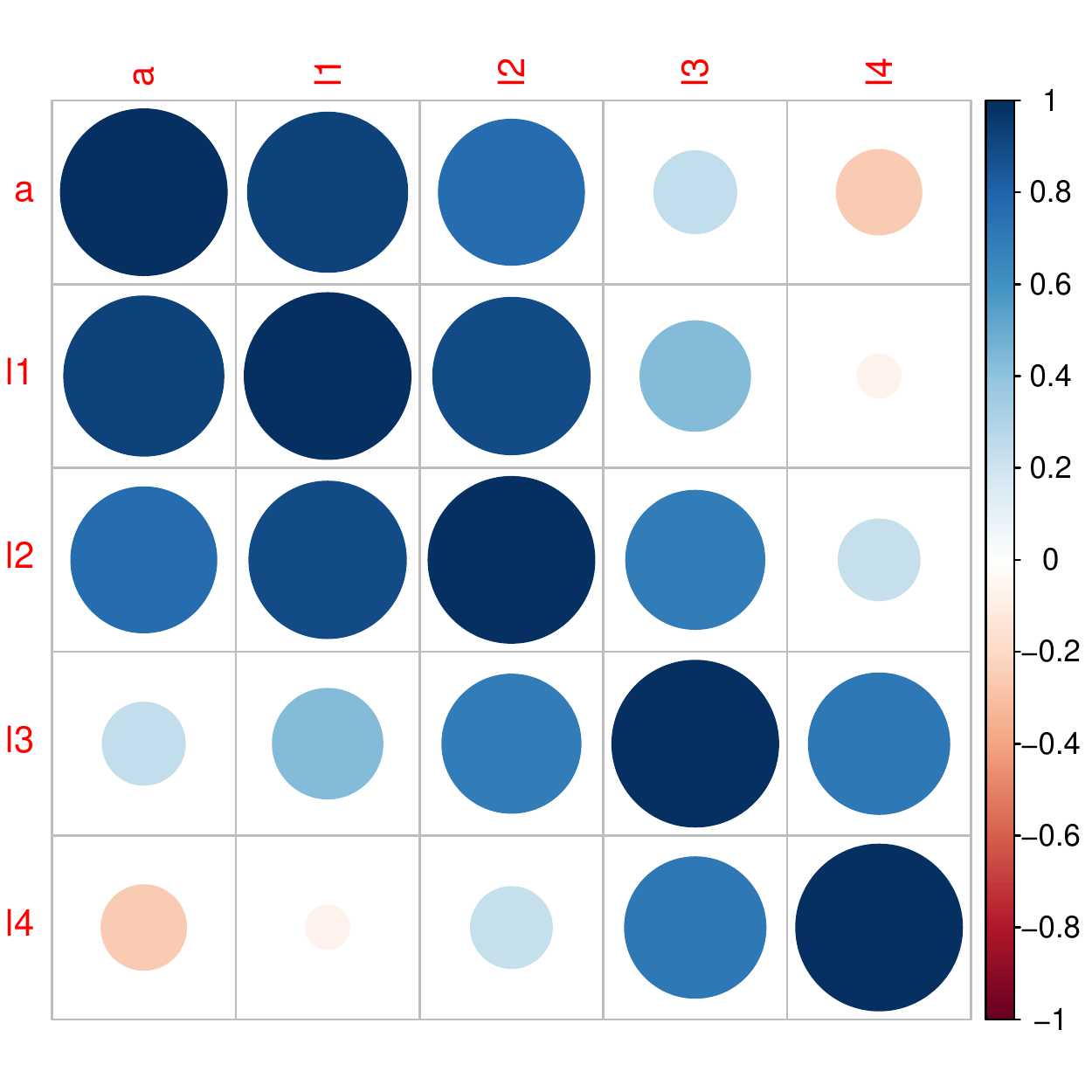} \qquad
    \includegraphics[width = .26\linewidth]{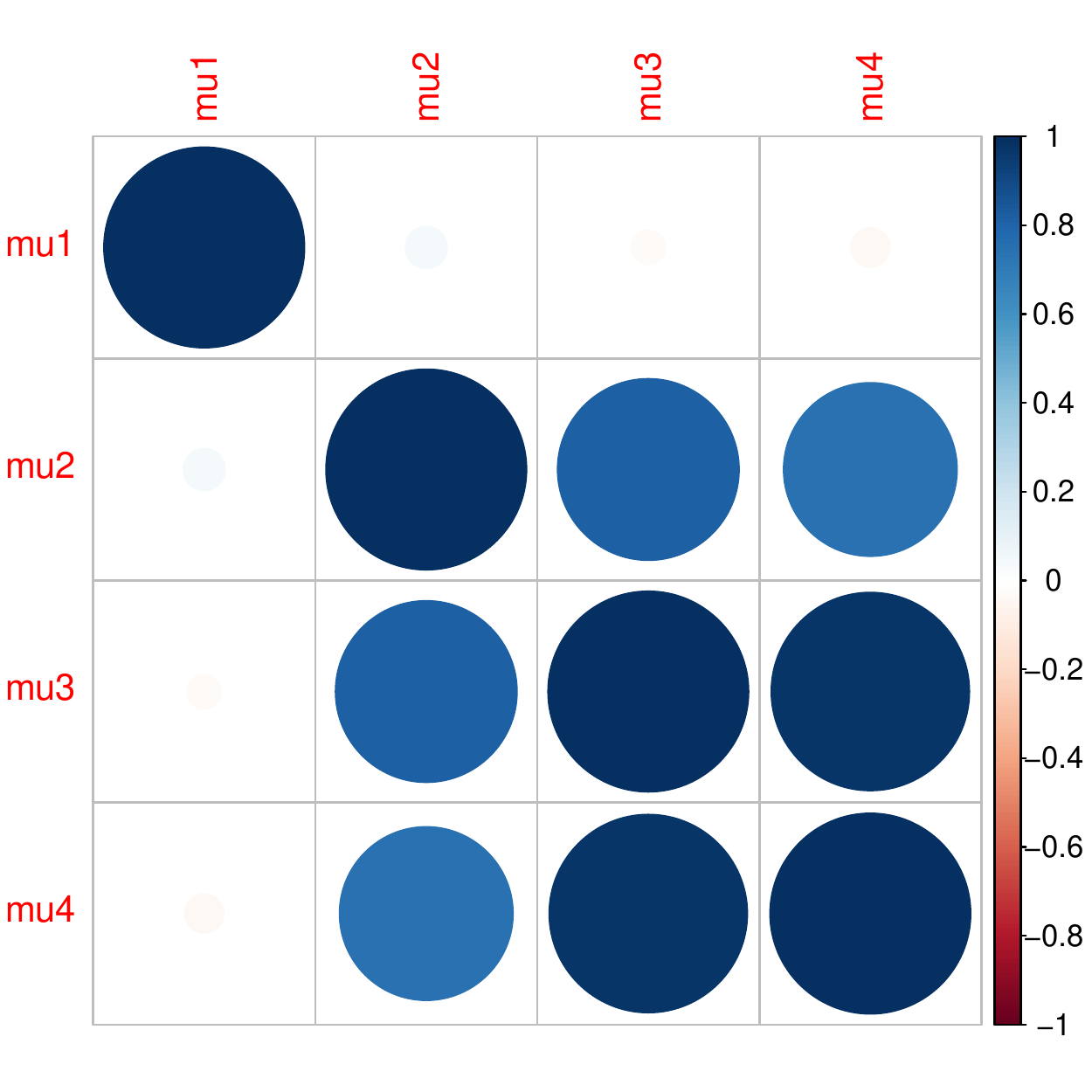}
    \caption{Estimated pairwise correlations for $\widetilde{\mu}_{h_1}$ (left) and $\widetilde{\mu}_{h_2}$ (right).}
    \label{fig:fish-corrplots}
\end{figure}
Using this proposal, the weight function is $g(\alpha, \lambda | \mathcal{D})$. Since the normalizing constants in the posterior distribution are unknown, only the SNIS estimator can be implemented. First, in order to visualize the complex correlation structures of $\widetilde{\mu}_{h_1}$ and $\widetilde{\mu}_{h_2}$, Figure~\ref{fig:fish-corrplots} plots the estimated sample correlation matrices of the corresponding $\widehat{\Omega}$. Since significant correlation between components for both $\widetilde{\mu}_{h_1}$ and $\widetilde{\mu}_{h_2}$ is evident, multivariate assessment of the quality of estimation is warranted.

\begin{figure}[htbp]
    \centering
    \includegraphics[width = .40\linewidth]{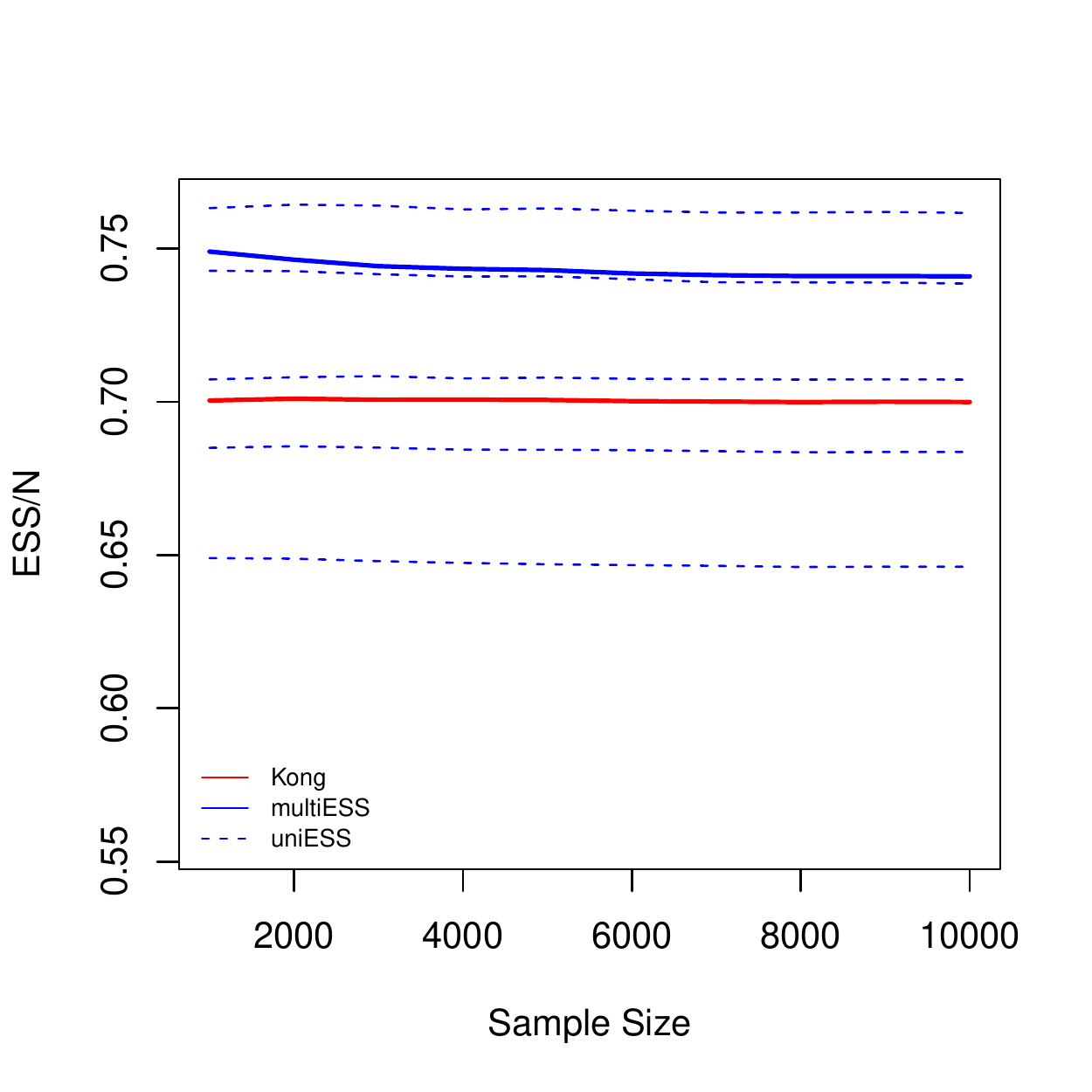} 
    \includegraphics[width = .40\linewidth]{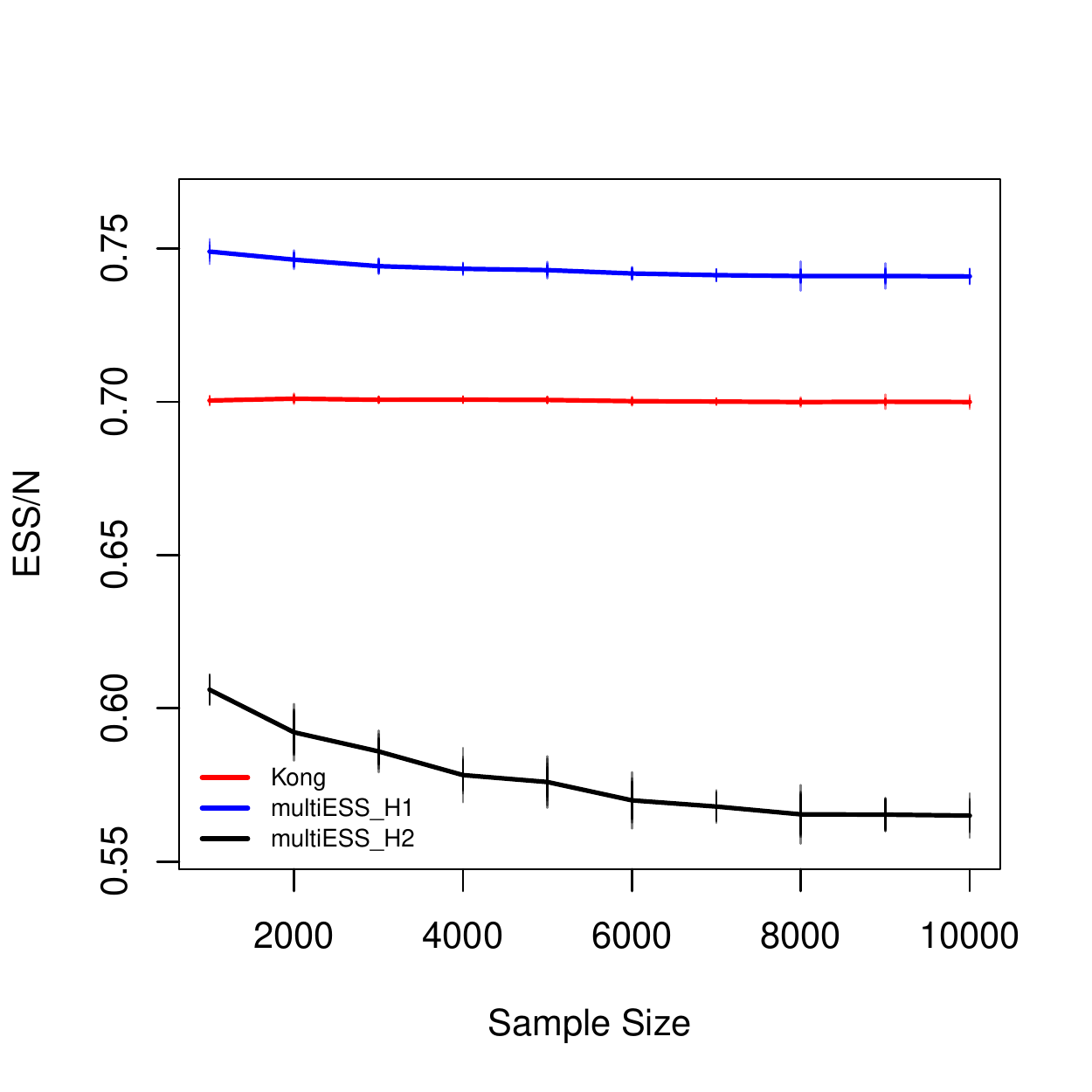}
    \caption{Left: ESS (Kong, univariate ESS, and M-ESS) vs sample size for $h(\theta) = h_1(\theta)$ where $\theta = (\alpha, \lambda_1, \dots, \lambda_{4})$. Right: ESS (Kong and M-ESS) vs sample size for $h_1(\theta)$ and $h_2(\theta)$. For each sample size, simulations are run for $500$ replications and the error bars for estimated ESS$/n$ are drawn using two standard deviations.}
    \label{fig:fish-ESSvsSampSize}
\end{figure}

The disparity between ESS estimation using mutivariate ESS and univariate ESS with five different components of $\theta$ can be seen in left side plot of Figure~\ref{fig:fish-ESSvsSampSize}. Here we present the estimated M-ESS for $h_1$, univariate ESS for each component of $h_1$, and $K_n$. Since interest is in estimating \textit{all} components of $h_1$, the individual univariate ESSs only provide partial information about the quality of estimation, whereas  M-ESS provides a complete understanding of the relative quality of estimation. 

In Figure~\ref{fig:fish-ESSvsSampSize} (right), $K_n$ is limited to only quantifying the quality of the proposal distribution in relation to $\pi$, and does not account for the different $h$ functions of interest. Further,  it is evident that for any same sample size $n$, the M-ESS for estimating $h_2$ is significantly smaller that the M-ESS for $h_1$. This indicates requirement of increased simulation effort for estimating $\mu_{h_2}$ and enables users to make informed decisions motivated by their primary objective. For instance, running the simulations with $\epsilon = 0.05$, we find that the necessary sample size (averaged over $10$ replications) using \eqref{eq:stop_practical} is $1.209 \times 10^4$ for estimating $\mu_{h_1}$ and $1.589 \times 10^4$ for estimating $\mu_{h_2}$.

\subsection{ {{Bayesian Lasso using INLA}}} \label{ex:lasso}
{{
\cite{berild2022isinla} propose an importance sampling based extension of INLA. INLA is an approximate Bayesian inference tool for models with a latent Gaussian structure. INLA focuses on quick approximate marginal posterior inference, as opposed to exact joint inference often done by MCMC. Although INLA is applicable to a relatively small class of models,  \cite{berild2022isinla} propose an importance sampling based method to expand this class to conditional latent Gaussian models.}}

{{
Let $y \in \mathbb{R}^m$ be the vector of observed $m$ responses and $X$ be an $m \times p$ model matrix made of $p$ regression covariates. For $\tau > 0$, consider the following Gaussian likelihood:
\[
y \sim  {\mathcal{N}}_m(X\beta, \tau^{-1} I_m)\,.
\]
For Bayesian lasso, a Laplace prior is put on $\beta$ and \cite{berild2022isinla,gomez2018markov} assume the default R-INLA prior for $\tau$: $\tau \sim \text{Gamma}(1, .00005)$. 
}}

{{
Since the Laplace prior is not available in R-INLA, \cite{gomez2018markov} propose a method that combines MCMC and INLA to fit this model by conditioning on $\beta$. Using this key idea, \cite{berild2022isinla} combine IS and INLA to fit this model. Using notation in \cite{berild2022isinla}, let $z = (\beta, \tau)$ with $z_c = \beta$ and $z_{-c} = \tau$; $z_c$ denotes the parameters being conditioned on and $z_{-c}$ denotes all other parameters. Given $z_c$, R-INLA provides numerical approximations to conditional posterior marginal $\pi(z_{-c}| y, z_c)$ and conditional marginal likelihood $\pi(y|z_c)$. In this situation, note that
\[
\pi(z_c | y) \propto \pi(y | z_c) \pi(z_c)\,,
\]
where $\pi(y | z_c)$ can be approximated by R-INLA. Therefore, the posterior expectation of $z_c$, denoted by $\mu_{z_c}:= \int z_c \pi(z_c | y) dz_c$, can be estimated using SNIS; $h(x)$ here is thus the identity function. The  samples $\{z_c^i\}_{i=1}^{n}$ are generated from a proposal density $g$ and the corresponding weights are
\[
w_i := \dfrac{\pi(y | z_c^i) \pi(z_c^i)}{g(z_c^i)}\,.
\]}}

{{
\citet[Section 4.2]{berild2022isinla} presents a detailed discussion on the need for an effective sample size that accounts for both the choice of $h(x)$ and does not require looking at many univariate ESSs. They acknowledge and indicate that one of the only possible quantities available is the modified ESS of \cite{Owen:2013}. Let $h_j(x)$ denote the $j$th component of the function $h$, then \cite{Owen:2013} proposes the following:
\[
\tilde{w}_i(h_j) := \dfrac{|h_j(X_i)| w_i}{ \sum_{t=1}^{n} |h_j(X_t)| w_t} \qquad\qquad \text{ and } \qquad \qquad \text{oESS}_j := \dfrac{1}{\sum_{t=1}^{n} \tilde{w}_t(h_j)^2}\,.
\]}}

{{
\cite{berild2022isinla} analyze the \texttt{Hitters} dataset available in \texttt{ISLR} library in \texttt{R}. Here the response is the salary variable and although they choose five numerical covariates, we analyze the complete dataset with all 19 covariates. Similar to \cite{berild2022isinla}, we use a  multivariate Student-$t$ proposal, $t_{\nu}(\mu_k, \Sigma_k)$ with $\nu = 3$; the mean $\mu_k$ and covariance $\Sigma_k$ are chosen via preliminary IS explorations.}}

\begin{figure}[h]
    \centering
    \includegraphics[width = 0.5\linewidth]{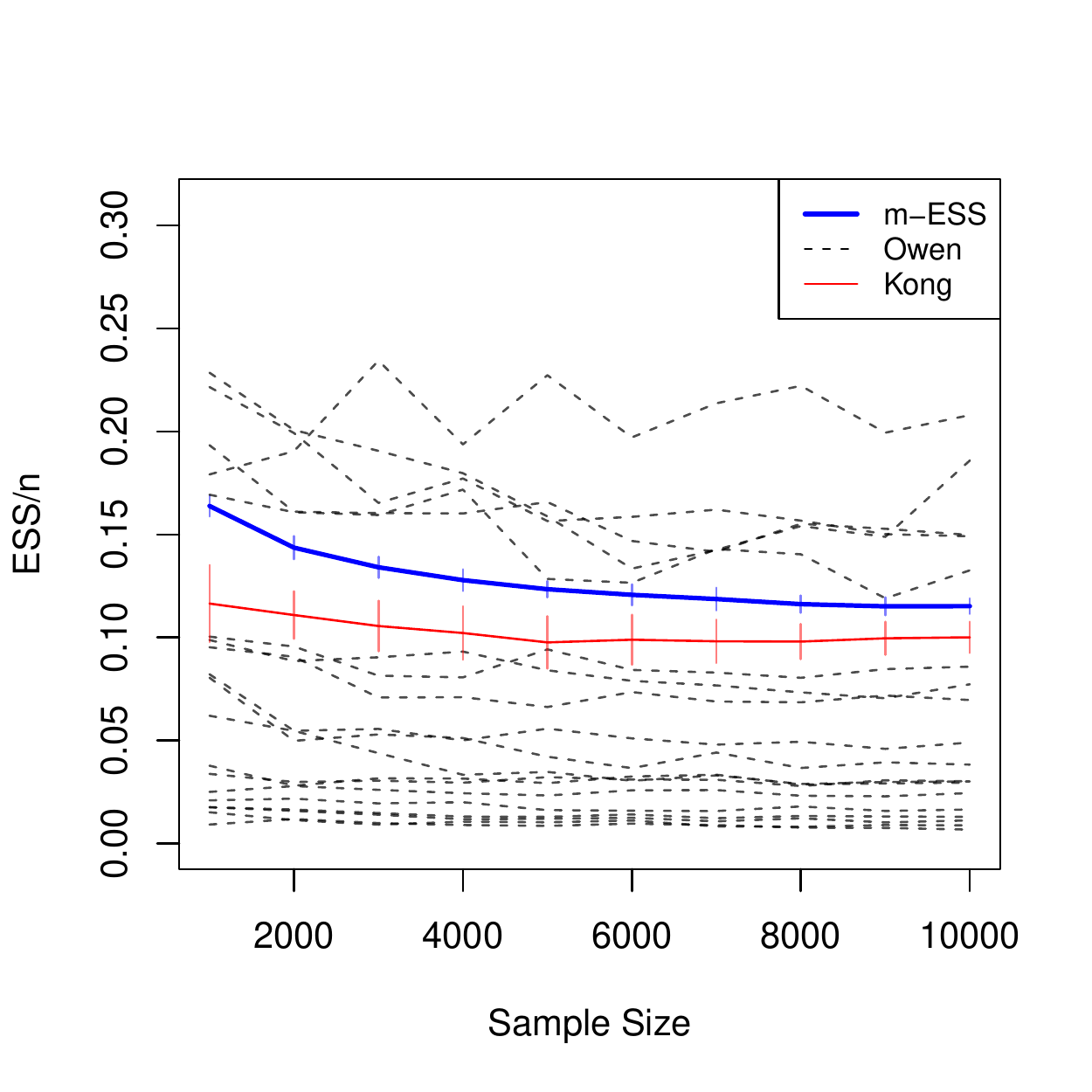}
    \caption{ {{Estimated ESS$/n$  vs $n$ (sample size), averaged over 10 replications.}}}
    \label{fig:lasso}
\end{figure}

{{
Figure~\ref{fig:lasso} presents ESS$/n$ versus Monte Carlo sample size $n$ for M-ESS, $K_n$, and oESS$_j$ (for each component), averaged over 10 replications. It is evident that the quality of estimation of the proposed M-ESS remains reasonably steady as a function of $n$ and there is significant difference between M-ESS and $K_n$. Further, oESS$_j$, although accounting for the choice of $h$, ignores the multivariate nature of the estimation problem and returns vastly differing estimation quality over the components.}}

{{
It is important here to also add that \cite{berild2022isinla} compare their IS based method to the MCMC based method of \cite{gomez2018markov}, by comparing the effective sample sizes of the two algorithms; \cite{berild2022isinla} use the ESSs of both \cite{kong1992note} and \cite{Owen:2013}. However, although still called ``effective sample size", both $K_n$ and oESS diverge significantly from the effective sample size used in MCMC. M-ESS, on the other hand, can directly be used to compare quality of estimation between IS and MCMC. In other words, if the M-ESS of an IS estimator is more than the effective sample size of an MCMC estimator, it can be correctly concluded that IS yields a more efficient estimator. The same conclusion cannot be made for $K_n$ and oESS.
}}
\section{Discussion}

We present a practical sequential stopping rule for IS, based on self-assessing the quality of estimation in a multivariate setting.  Note that choosing and adapting a proposal distribution is a rich and important area of work, known as adaptive importance sampling (AIS), and many existing algorithms are devoted to this task, e.g., PMC \citep{Cappe04}, AMIS \citep{CORNUET12}, M-PMC \citep{Cappe08}, LAIS \citep{martino2017layered}, DM-PMC \citep{elvira2017improving}, or O-PMC \citep{elvira2021optimized}  \citep[see][for a detailed review]{bugallo2017adaptive}. These existing methods do not consider an adaptive choice of the number of proposals nor the number of samples. We are confident that the proposed methodology can be useful to develop more efficient AIS algorithms. %

Traditionally, ESS is defined as the relative variance of the vanilla Monte Carlo estimator and the chosen estimator (see \cite{kong1992note} and the discussion in \cite{elvira2018rethinking}). 
One can argue against the choice of the vanilla Monte Carlo estimator as the baseline; this may be seen in applications to rare event simulation \citep{owen2019importance}, and therefore, the ESS would not be as informative in such scenarios. Our proposed stopping rule in \eqref{eq:stop_practical} and Theorem~\ref{th:ssr} can be adapted to such changes in the  baseline. 

{{As highlighted in the example in Section~\ref{ex:weibull}, the stopping rule and quality of estimation critically depends on the function of interest $h$. It is then imperative that users first decide their $h$ of interest, before commencing the IS procedure. In scenarios where a specific $h$ may not be of interest, the $K_n$ metric of \cite{kong1992note} would be useful.}} While $K_n$ does not translate into the number of effective samples from the target, it can still be useful in understanding the quality of the weights.  Thus, $K_n$ can be used as a qualifier to ascertain the quality of the proposal distribution for a particular target.

Other IS estimators have been proposed that promise improvement in estimation quality \citep[see][]{martino2018comparison, elvira2019generalized, vehtari2015pareto} by utilising weight modification techniques. These methods exhibit variance reduction in theory, but there are no known methods of estimating their variance other than bootstrap. Thus, variance estimation of other IS estimators is an interesting and critical line of future research. 

\section{Acknowledgements}

The authors thank the anonymous referee for their useful suggestions that significantly improved the paper. The work of D.V. is supported by SERB (SPG/2021/001322). The work of V. E. is supported by the \emph{Agence Nationale de la Recherche} of France under PISCES (ANR-17-CE40-0031-01) and the Leverhulme Research Fellowship (RF-2021-593).

\appendix
\section{Appendix}

\subsection{Variance of the SNIS estimator} \label{appendix:variance_of_SNIS}

{Versions of the proof of the asymptotic normality of $\tilde{\mu}_h$ is available in \cite{rice2006mathematical,nilakanta2020output}, however, we present the proof here for completeness.} Define the mapping $k: \mathbb{R}^s \to \mathbb{R}^{p+1}$ such that $X \mapsto (h(X)w(X),\, w(X)).$ If $\{X_i\}_{i=1}^{n}$ are i.i.d. samples from the proposal, this allows us to define iid random variables $Y_i := (h(X_i)w_i,\, w_i)$, for $i = 1, \dots, n$ so that $\theta :=\mathbb{E}_{q}[Y_1] =  (Z\mu_h, Z)^T$ and let $\Lambda:=\Var_{q}(Y_1)$ denote the $(p+1) \times (p+1)$ covariance matrix of $Y_1$ and we assume that $\Lambda < \infty$.  Let $\bar{Y}_n$ denote the sample average of $\{Y_i\}_{i=1}^n$. By a standard central limit theorem,
\[
\sqrt{n}(\bar{Y}_n - \theta) \xrightarrow[]{d} {\mathcal{N}}(0, \Lambda)\qquad \text{ as } n \to \infty\,.
\]
Define the function $g: \mathbb{R}^{p} \times \mathbb{R} \to \mathbb{R}^p$  as $g(a,b) = a/b$ for $a \in \mathbb{R}^p \text{ and } b \in \mathbb{R}$.  Then using the multivariate delta method \citep[see][]{lehmann2004elements} for variance of ratio of means and a second-order approximation,
 \[
 \sqrt{n}(g(\bar{Y}_n) - g(\theta)) \xrightarrow[]{d} {\mathcal{N}}_p(0, \nabla g(\theta)^T \Lambda\nabla  g(\theta)) \qquad \text{ as } n \to \infty \,,
 \]
 where
\[
\nabla g(\theta) = \begin{pmatrix}
\dfrac{I_p}{Z} &\,\,\,\, -\dfrac{\mu_h}{Z^2}
\end{pmatrix}^T\,.
\]
Note that $g$ translates the sample average $\bar{Y}_n$ to the SNIS estimator as
\[
g(\bar{Y}_n) = g \left(\dfrac{\sum_{i=1}^{n} h(X_i)w_i}{
n}, \dfrac{\sum_{i=1}^{n} w_i}{n}\right)  = \dfrac{\sum_{i=1}^{n} h(X_i)w_i}{\sum_{i=1}^{n} w_i} = \widetilde{\mu}_h\,.
\]
Therefore, as $n \to \infty$, we have 
\begin{equation} \label{eq:asymp_law_of_SNIS}
    \sqrt{n}(\widetilde{\mu}_h - \mu_h) \xrightarrow[]{d} {\mathcal{N}}_p(0, {\Omega}),
\end{equation}
{with}
 \[
 \Omega = \nabla g(\theta)^T \Lambda\nabla g(\theta) = \dfrac{\mathbb{E}_q[w(X)^2(h(X) - \mu)(h(X) - \mu)^T]}{\mathbb{E}_q[w(X)]^2}.
 \]

\subsection{Proof of Theorem~\ref{th:ssr}}

\begin{proof}

For the purpose of this proof, we  extend the notation of $\widetilde{\mu}_h \text{ and } \widehat{\Omega}$ to explicitly display the number of samples used in estimation. Let $\widetilde{\mu}_h(m) \text{ and } \widehat{\Omega}(m)$ denote the estimators $\widetilde{\mu}_h \text{ and } \widehat{\Omega}$ constructed using $m$ importance samples. First, we show that as $\epsilon \to 0$, $T^*(\epsilon) \to \infty$. Consider $t^*(\epsilon) = \inf \left\{n \geq 0: s(n) \leq \epsilon R_n(X) \right\}$. As $\epsilon \to 0$, $t^*(\epsilon) \to \infty$ and $T^*(\epsilon) > t^*(\epsilon)$, yielding $T^*(\epsilon) \to \infty$.

Define $V(n) = \text{Vol}({C_{\alpha}(n)})^{1/p} + s(n)$ and
\[
d_{\alpha, p} :=  \dfrac{2 \pi^{p/2}}{p \Gamma(p/2)}(\chi^2_{1-\alpha, p})^{p/2}\,.
\]
Recall from \eqref{eq:confidence_region} and the statement of Theorem~\ref{th:ssr}
\[
\text{Vol}({C_{\alpha}(n)}) = \dfrac{2 \pi^{p/2}}{p \Gamma(p/2)} \left(\dfrac{\chi^2_{1-\alpha, p}}{n}\right)^{p/2} |\widehat{\Omega}|^{1/2} \; \text{ and } \; s(n) = \epsilon R_n(X)I(n < n^*) + n^{-1}\,.
\]
Using the fact that $s(n) = o(n^{-1/2})$ and $\widehat{\Omega}$ is consistent, we have $n^{1/2}V(n) \to (d_{\alpha,p}|\Omega|^{1/2})^{1/p}$ as $n \to \infty$. The following limit follows directly from \cite{vats2017multivariate}
\[
\lim_{\epsilon \to 0} \epsilon T^*(\epsilon)^{1/2} = d_{\alpha, p}^{1/p}\dfrac{\abs{\Omega}^{1/2p}}{R(X)}\,.
\]
Using the above, a functional delta method, and a standard random time change argument \citep[][ p. 144]{billingsley2013convergence} we have,
\[
\sqrt{T^*(\epsilon)} \left[ \widehat{\Omega}(T^*(\epsilon)) \right]^{-1/2} \left(\widetilde{\mu}_h(T^*(\epsilon)) - \mu_h \right) \xrightarrow[]{d} {\mathcal{N}}_p(0, I_p)\,.
\]
As a consequence, as $\epsilon \to 0$, with probability, $\text{Pr}[\mu_h \in C_{\alpha}\left( T^*(\epsilon)\right)] \to 1 - \alpha$.
%
\end{proof}

\subsection{True Variances for Multivariate Gaussian}

Recall from Eq.\eqref{eq:SNIS_variance}, the limiting variance of $\widetilde{\mu}_h$ is given by
\[
\Omega = \int_{\mathbb{R}^p} \dfrac{(h(x) - \mu_h)(h(x) - \mu_h)^T \pi(x)^2}{q(x)} dx\,.
\]
In Section~\ref{ex:normal}, $h(x) = x,\, \pi = N(\mu, \Lambda), \text{ and } q = N(\mu, \Upsilon)$. Plugging this in \eqref{eq:SNIS_variance}, we get
\begin{align*} 
    \Omega &= \int_{\mathbb{R}^p} \dfrac{(x - \mu)( x - \mu)^T \pi(x)^2}{q(x)} dx \nonumber \\
    &= \dfrac{|\Upsilon|^{1/2}}{|\Lambda|}\int_{\mathbb{R}^p} \dfrac{(x - \mu)( x - \mu)^T}{(2 \pi)^{p/2}} \exp \left\{-(x-\mu)^T\Lambda^{-1}(x - \mu) + \dfrac{(x-\mu)^T\Upsilon^{-1} (x-\mu)}{2 } \right\} dx \nonumber \\
    &= \dfrac{|\Upsilon|^{1/2}}{|\Lambda| |2\Lambda^{-1} - \Upsilon^{-1}|^{1/2}} (2\Lambda^{-1} - \Upsilon^{-1})^{-1}\,.
\end{align*}

Next,
\begin{align} \label{eq:normal-trueSigma}
    \Omega_U &=\int_{\mathbb{R}^p} \dfrac{(h(x)h(x)^T \pi(x)^2}{q(x)} dx - \mu\mu^T\\
    & = \int_{\mathbb{R}^p} \dfrac{xx^T \pi(x)^2}{q(x)} dx - \mu \mu^T\nonumber \\
    &= \dfrac{|\Upsilon|^{1/2}}{|\Lambda|}\int_{\mathbb{R}^p} \dfrac{xx^T}{(2 \pi)^{p/2}} \exp \left\{-(x-\mu)^T\Lambda^{-1}(x - \mu) + \dfrac{(x-\mu)^T\Upsilon^{-1} (x-\mu)}{2 } \right\} dx - \mu \mu^T \nonumber \\
    &= \dfrac{|\Upsilon|^{1/2}}{|\Lambda| |2\Lambda^{-1} - \Upsilon^{-1}|^{1/2}} \left[(2\Lambda^{-1} - \Upsilon^{-1})^{-1} + \mu \mu^T\right] - \mu \mu^T\,.
\end{align}

\subsection{Details of Bayesian Multi-Step Step-Stress Model}

The combined likelihood for the fish dataset in Section~\ref{ex:weibull} with $\theta = (\alpha, \lambda_1, \lambda_2, \lambda_3, \lambda_4)$ is, 
\[
    L(\theta \mid \mathcal{D}) \propto \alpha^r \lambda_1^n \lambda_2^n \lambda_{3}^n \lambda_{4}^n \left( \prod_{i=1}^{n} t_i^{\alpha - 1} \right) e^{-\left(\lambda_1 D_1(\alpha) +  \lambda_{3} + \lambda_{4} D_{4}(\alpha) \right)}\,,
\]
where
\begin{equation*}
   D_j(\alpha) = \left( \sum_{i = \bar{n}_{j-1}-1}^{\bar{n}_j} t_i^{\alpha} + (n - \bar{n}_j)\tau_j^\alpha - (n - \bar{n}_{j-1}) \tau_{j-1}^\alpha \right), \qquad \bar{n}_0 = \tau_0 = 0,\,\, \bar{n}_{4} = n\,. 
\end{equation*}

Let $S(\alpha) = \min\left\{D_1(\alpha), D_2(\alpha), D_3(\alpha), D_{4}(\alpha)\right\}$ and $J = \min(n_1, n_2, n_3, n_{4})$, then $\pi_1^*(\lambda | \alpha, \mathcal{D})$, $\pi^\ast_2(\alpha | \mathcal{D})$, and $g(\alpha, \lambda | \mathcal{D})$ from the full posterior distribution in Eq.\eqref{eq:weibull-posterior} are 
\begin{align*}
    \pi^\ast_1(\lambda | \alpha, \mathcal{D}) & = \text{density of } \text{ODG}(a_o + 4J, b_o    + S(\alpha), a_1 + J, a_2 + J, a_3 + J, a_{4}+J)\\
    \pi^\ast_2(\alpha | \mathcal{D}) &  = \text{density of } \text{Gamma} \left(n + a, b - \sum_{i=1}^{n}\ln t_i \right)\\
    g(\alpha, \lambda | \mathcal{D}) &= \dfrac{\prod_{j=1}^{4} \left[\lambda_j^{n_j - J} e^{-\lambda_j(D_j(\alpha) - S(\alpha))} \right]}{[b_o + S(\alpha)]^{a_o + 4J}}\,. 
\end{align*}

\bibliographystyle{apalike}
\bibliography{sample}

\end{document}